# Ultraviolet Exciton-Polaritons in Silver Phenylthiolate

Bongjun Choi[1], Bonnie Chen[2], Thuc T. Mai[3], Rahul Rao[3], Adam D. Alfieri[1], Du Chen[4,5], Peijun Guo[4,5], Ha-Reem Kim[1,6,7], Michael A. Altvater[3], Nicholas A. Glavin[3], Deep Jariwala[1,2*]

[1]Department of Electrical and Systems Engineering, University of Pennsylvania, Philadelphia, Pennsylvania 19104, United States

[2]Department of Materials Science and Engineering, University of Pennsylvania, Philadelphia, Pennsylvania 19104, United States

[3]Materials and Manufacturing Directorate, Air Force Research Laboratory, Wright-Patterson Air Force Base, OH 45433, USA

[4]Department of Chemical & Environmental Engineering, Yale University, New Haven, CT 06520, USA

[5]Energy Sciences Institute, Yale University, West Haven, CT 06516, USA

[6]Department of Physics, The Catholic University of Korea, Bucheon 14662, Republic of Korea

[7]Department of Physics and Astronomy, and Institute of Applied Physics, Seoul National University, Seoul 08826, Republic of Korea

[*] Corresponding authors: dmj@seas.upenn.edu

## Abstract

Ultraviolet (UV) exciton-polaritons (EPs) enable nonlinear optics, polaritonic lasing, and polariton-mediated photochemistry in the short-wavelength regime, yet progress has been limited on this topic due to the scarcity of materials that combine large oscillator strength with stable and narrow UV excitons. Here, we demonstrate UV EPs in silver phenylthiolate (Thiorene, AgSPh), a van der Waals (vdW) layered metal-organic chalcogenolate (MOC) that forms a natural multi-quantum-well (MQW) architecture showing strong excitonic features. Imaging spectroscopic ellipsometry reveals a pronounced in-plane excitonic resonance at 3.46 eV with a narrow linewidth of ~60 meV, strong UV birefringence ($\Delta n \approx 0.3$), and a high refractive index ($n \approx 2.1$). Temperature-dependent photoluminescence (PL) shows excitonic emission with a large Stokes shift and substantial exciton-phonon coupling. In both open (self-cavity) and closed cavities, thickness-dependent and angle-resolved reflectance spectra exhibit clear anticrossing, yielding large Rabi splittings of approximately 500 meV. These values are among the largest reported in the UV, positioning thiorene as a promising platform for UV polariton lasers and polariton-enabled photochemistry.

## Introduction

Exciton-polaritons (EPs) are hybrid light-matter quasiparticles formed when an excitonic transition strongly couples to a confined photonic mode, giving rise to new eigenstates separated by the Rabi splitting[1,2]. Owing to their mixed photonic and excitonic character—combining a light effective mass with strong Coulomb interactions—EPs enable rich collective and nonlinear phenomena such as Bose-Einstein condensation[3,4], low-threshold polaritonic lasing[5,6], and polariton-enabled photochemistry[2,7]. In all these applications, a large Rabi splitting is desirable: it provides clear spectral separation of the polariton branches, enabling operation deep within the strong-coupling regime [2].

Because light-matter coupling strength scales with both the exciton oscillator strength and the electromagnetic field confinement[1], excitonic materials with large refractive indices operating in the ultraviolet (UV) are promising platforms for strong coupling. In suitable material systems with sufficiently strong and stable UV excitonic resonances, large Rabi splitting can be achieved, enabling engineering of optical dispersion at technologically relevant UV wavelengths. However, UV EPs remain constrained by the scarcity of materials that simultaneously offer (i) large exciton binding energy, (ii) large oscillator strength, and (iii) intrinsically narrow linewidths. Conventional UV excitonic platforms such as zinc oxide (ZnO) and gallium nitride (GaN) exhibit room-temperature strong coupling, with Rabi splitting of 50-80 meV for ZnO[8,9] and 64-162 meV for GaN[10,11], yet their bulk exciton binding energies (~60 meV for ZnO[12] and ~26 meV for GaN[13]) are relatively modest. Even in wider-bandgap aluminum nitride (AlN), experimentally demonstrated Rabi splitting remains limited to 44 meV, partly due to the lack of a strong excitonic response in the UV regime[14]. Consequently, materials capable of sustaining sharply defined UV excitons and large Rabi splitting remain scarce.

Metal-organic chalcogenolates (MOCs) have recently emerged as a promising class of low-dimensional semiconductors with an intrinsic multi-quantum well (MQW) structure[15-17]. Their alternating inorganic-organic layered structure introduces strong dielectric confinement, suppressing Coulomb screening and yielding tightly bound excitons[15]. Within the silver phenylchalcogenolate family (AgEPh, E= S, Se, and Te), silver phenylselenolate (AgSePh, hereafter mithrene) has demonstrated remarkable excitonic responses and ultrastrong light-matter interaction in the blue regime, establishing MOC as a potent platform for EPs[17-19]. Silver phenylthiolate (AgSPh, hereafter thiorene), possessing a wider bandgap than its selenide analogue, is predicted by density functional theory (DFT, GW-Bethe-Salpeter equation (BSE)) calculations to host an exceptionally large exciton binding energy (~479 meV) due to its MQW structure, and it exhibits strong UV excitonic absorption[20,21]. Such a large binding energy suggests stable UV excitons even at room-temperature–an essential prerequisite for achieving large Rabi splitting and stable EPs in the UV. However, systematic exploration of thiorene's optical properties and polaritonic behavior of thiorene remains limited.

Here, we establish thiorene as a UV strong-coupling platform through combined structural, dielectric function, and polaritonic characterization. Imaging spectroscopic ellipsometry reveals a sharp in-plane lowest exciton at 3.46 eV with a narrow linewidth (~60 meV), while the out-of-plane response remains featureless due to weak interlayer electron wavefunction overlap, similar to mithrene[17]. The resulting anisotropic dielectric response produces giant UV birefringence (Δn=~0.3). Photoluminescence (PL) spectroscopy reveals excitonic emission characterized by a pronounced blue shift and linewidth narrowing with decreasing temperature, consistent with excitonic behavior[22], along with pronounced exciton-phonon coupling. Building on these properties, we demonstrate strong exciton-photon hybridization in both open (self-cavity) and closed cavity geometries, observing clear anticrossing behavior with large Rabi splitting of 488 meV and 512 meV, respectively–among the largest in the UV regime. The relatively high refractive index of thiorene (~2.1) enables self-hybridized EPs even in a simple open cavity configuration. The larger splitting in the closed cavity arises from reduced mode volume and enhanced confinement. Together, these results position thiorene as a compelling material platform for UV polariton lasers, dispersion-engineered photonic devices, and polariton-enabled photochemistry.

**Results and Discussion**

To investigate the excitonic properties of thiorene, we begin by synthesizing large single-crystalline thiorene via an organic single-phase reaction[23]. Thiorene crystallizes in the monoclinic Cc or $P2_1$ space group and forms a van der Waals (vdW) layered hybrid semiconductor (Fig. 1a) in which covalently connected Ag-S inorganic networks create electronically active sheets separated by organic phenyl spacers, analogous to the structural motif observed in mithrene[20,24]. The alternating inorganic/organic stacking forms a natural MQW with strong dielectric contrast. Dielectric confinement from the low-permittivity organic layers suppresses Coulomb screening, producing tightly bound excitons with large oscillator strength—advantageous for strong light-matter interactions[25]. Figure 1b shows a schematic illustration of an organic single-phase reaction, in which silver nitrate ($AgNO_3$) reacts with diphenyl disulfide ($Ph_2S_2$) in a mixed toluene/propylamine ($PrNH_2$) solvent. Propylamine can coordinate $Ag^+$ and modify reaction kinetics, a mechanism previously shown to enhance size and optical quality in mithrene by slowing conversion and suppressing defect formation during crystallization[23]. Our optimized growth yields single-crystalline thiorene with lateral dimensions exceeding 1 cm (Fig. 1c and supporting information (SI) Fig. S1). These large crystals are particularly valuable for polariton experiments because they reduce the impact of grain boundaries and allow optical probing on uniform single-crystal regions. Consistent with its vdW stacking, thiorene can be mechanically exfoliated into thin flakes with a smooth surface (Fig. 1c), confirmed by atomic force microscopy (AFM), as shown in SI Fig. S2. Thickness-controlled flakes obtained via mechanical exfoliation provide a practical route for engineering cavity length and optical field confinement in UV exciton-polaritonic structures[17,18] (e.g., open and closed cavity), as discussed further below.

Scanning electron microscopy (SEM) and energy-dispersive X-Ray spectroscopy (EDX) elemental mapping show spatially uniform Ag, S, and C signals across the examined crystal region, indicating macroscopic compositional homogeneity and the absence of obvious phase segregation or residual precursor-rich domains (Fig. 1d). Such uniformity is important for EPs hybridization because compositional gradients and impurity-rich regions can introduce localized electronic states that broaden excitonic resonances. Furthermore, the EDX spectrum acquired from the selected region of interest reveals dominant Ag and S peaks together with the expected C signal from the phenyl ring, aligned with the previous observation in the poly-crystalline thiorene film[26] (SI Fig. S3). Raman spectroscopy provides an additional fingerprint of the phenylthiolate framework, showing low-energy phonon modes (<200 $cm^{-1}$) associated with metal-chalcogen coordination in the inorganic quantum-well planes, along with phenyl ligand vibrations near 1000 cm$^{-1}$ as shown in Fig. 1e[27-30]. Comparison with mithrene indicates that the chalcogen-related modes exhibit a systematic shift upon S/Se substitution, whereas the phenyl-dominated modes remain largely unchanged (SI Fig. S4). Powder X-ray diffraction (PXRD) further confirms high crystallinity and the expected layered phase (Fig. 1f), consistent with previously reported structures of silver phenylthiolate[21,26,31]. The sharp low-angle Bragg reflections reveal a well-defined interlayer spacing, consistent with the vdW layered structure that forms an intrinsic MQW architecture[32]. Taken together, SEM/EDX, Raman, and PXRD confirm that the centimeter-scale crystals preserve the intended vdW layered architecture, ensuring the structural and chemical quality required for reliable extraction of dielectric functions and light-matter interaction studies.

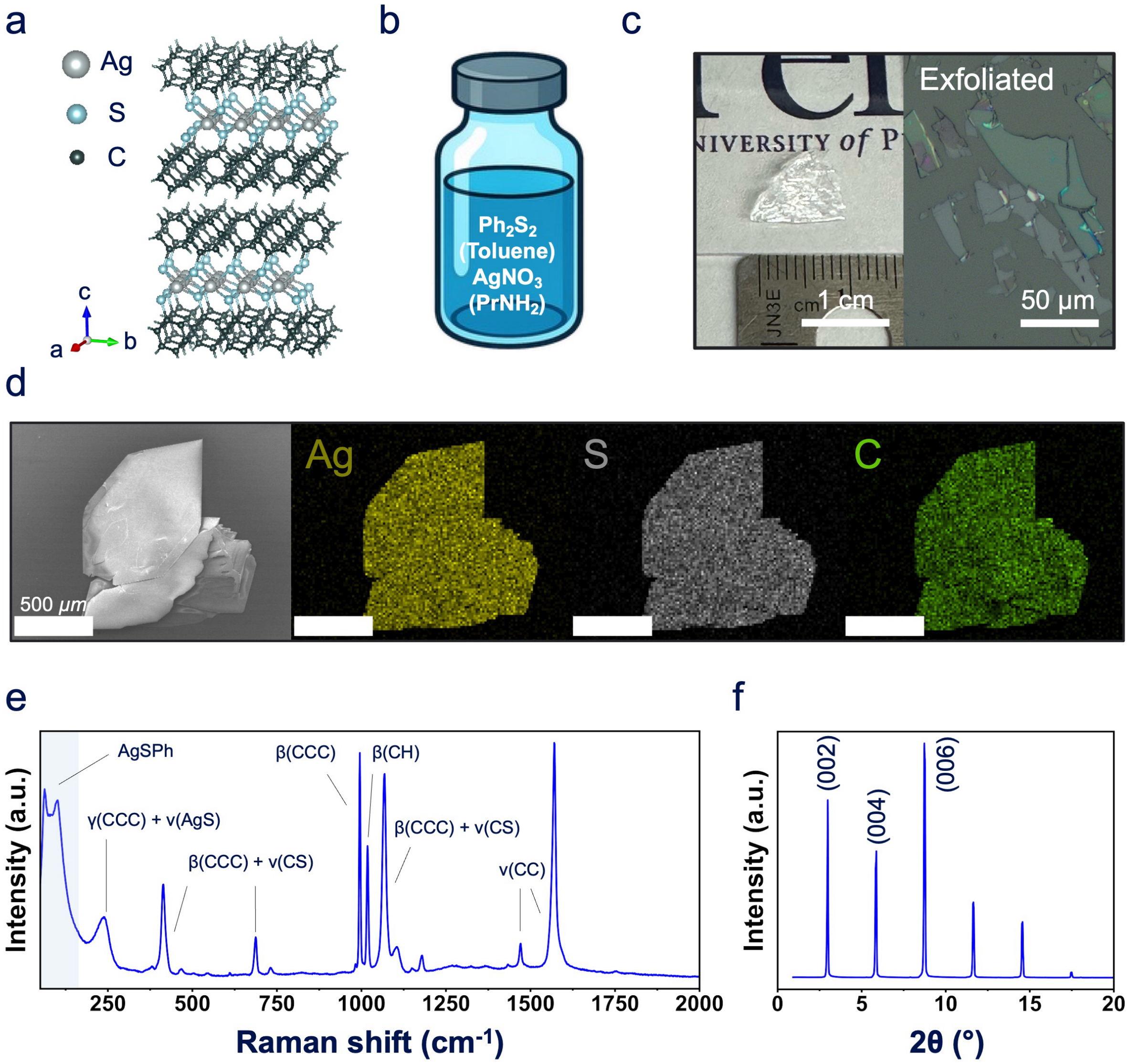

**Figure 1. Synthesis and structural characterization of thiorene (AgSPh).** (a) Layered crystal structure of thiorene, highlighting alternating inorganic Ag-S sheets and organic phenyl spacers, forming a natural multi-quantum well (MQW) structure. (b) Schematic illustration of amine-assisted solution synthesis of thiorene using silver nitrate ($AgNO_3$) and diphenyl disulfide ($Ph_2S_2$) in a toluene/propylamine($PrNH_2$) solution, enabling the large crystal synthesis essential for fundamental studies. (c) Photograph of centimeter-scale thiorene crystals and optical microscope image of mechanically exfoliated thiorene. (d) Scanning electron

microscopy (SEM) image and energy-dispersive X-Ray Spectroscopy (EDX) elemental maps (Ag, S, and C) showing spatially uniform composition across the flake. All scale bars indicate 500 μm. (e) Raman spectrum at room-temperature with characteristic vibrational modes of the phenylthiolate framework, including both inorganic (Ag-S coordination) and organic (phenyl-ring) modes. (f) Powder X-ray diffraction (PXRD) pattern at room-temperature confirming the crystalline layered phase of the synthesized thiorene.

To quantitatively probe the excitonic response of thiorene, we investigate its anisotropic optical constants via imaging spectroscopic ellipsometry, enabling the micro-scale mapping of ellipsometry parameters (Ψ and Δ). Spectroscopic ellipsometry is a powerful technique for characterizing optical properties, as it sensitively captures both the amplitude (Ψ) and phase (Δ) changes of reflected light[33]. Thiorene exhibits pronounced optical anisotropy (in-plane vs. out-of-plane) owing to its intrinsically layered structure[20]. We perform measurements at multiple angles of incidence to reliably capture the excitonic response from 250 nm to 1000 nm. The mapping of ellipsometry parameters (Ψ and Δ) reveals highly uniform reflection across the flake, indicative of the high optical quality of thiorene (SI Fig. S5a). The fitted results using the uniaxial model show excellent agreement with the experimental data (Ψ and Δ, SI Fig. S5b). The resulting complex refractive indices ($\tilde{n} = n + ik$) reveal a strongly direction-dependent response (in-plane (∥) vs. out-of-plane (⊥)) (Fig. 2a and Table 1). The in-plane dielectric function is well described by a multi-oscillator Lorentz model with resonances centered at 3.46 eV, 3.56 eV, and 4.50 eV. Furthermore, the lowest-energy resonance (3.46 eV) exhibits a larger extinction coefficient ($k$) and a narrow linewidth (~60 meV) than the higher-energy feature at 4.50 eV, which is characterized by a broader linewidth. These strong and multiple excitonic transitions are consistent with recent first-principles DFT predictions and are highly favorable for realizing robust light-matter coupling[1,20]. The overall agreement supports attributing the sharp UV features in the dielectric response to intrinsic excitonic resonances rather than to defect-related absorption. In contrast, the out-of-plane dielectric response shows no discernible resonance in this spectral window since the electronic wavefunction overlap along the out-of-plane direction is minimal due to the organic spacer, similar to the behavior observed in mithrene. This behavior is well modelled by the Cauchy model[17]. Moreover, thiorene exhibits a relatively high refractive index of ~2.1, which is favorable for sustaining optical self-cavity modes. This high index further enables self-hybridized EPs, as discussed later. We validate the ellipsometry-derived complex dielectric functions using reflectance spectroscopy as a function of the incidence angle (Fig. 2b). The reflectance spectra show excellent agreement between experiment and transfer matrix method (TMM) calculations at multiple angles of incidence, confirming the reliability of the extracted refractive indices in Fig. 2a.

From the anisotropic optical constants, we compute the birefringence magnitude $|\Delta n|_{\perp,\parallel} = \left|n_{\parallel} - n_{\perp}\right| \, or \, \left|n_{\parallel,a} - n_{\parallel,b}\right|$ (Fig. 2c). To contextualize this exciton-driven anisotropy, we systematically benchmark the absolute birefringence ($|\Delta n|$) of thiorene with well-established birefringent materials, including barium titanate ($BaTiO_3$), barium titanium sulfide ($BaTiS_3$), and calcite ($CaCO_3$), as well as strongly anisotropic layered semiconductors such as hexagonal boron nitride (hBN), black phosphorus (BP), rhenium disulfide ($ReS_2$), niobium oxychloride ($NbOCl_2$), and mithrene, using reported experimental and calculated refractive indices along in- and out-of-plane directions[17,34-40]. Thiorene shows substantial birefringence in the UV regime due to the excitonic resonance, reaching values on the order of typical giant birefringence (e.g., $\Delta n \geq 0.3$). Across the broader UV range (λ = 250-400 nm), the

birefringence of thiorene averages approximately 0.26. For the comparison, classical birefringent crystals such as $CaCO_3$ exhibit a smaller average birefringence (~0.20) in this spectral range. In contrast, vdW and layered materials can reach larger values, including hBN (~0.67) and $BaTiS_3$ (~0.48), while $NbOCl_2$ exhibits an exceptionally large birefringence exceeding 1.15 on average within the UV regime. The birefringence of thiorene is comparable to that of conventional crystals and falls within the range observed for emerging anisotropic materials, suggesting its potential utility in UV photonics.

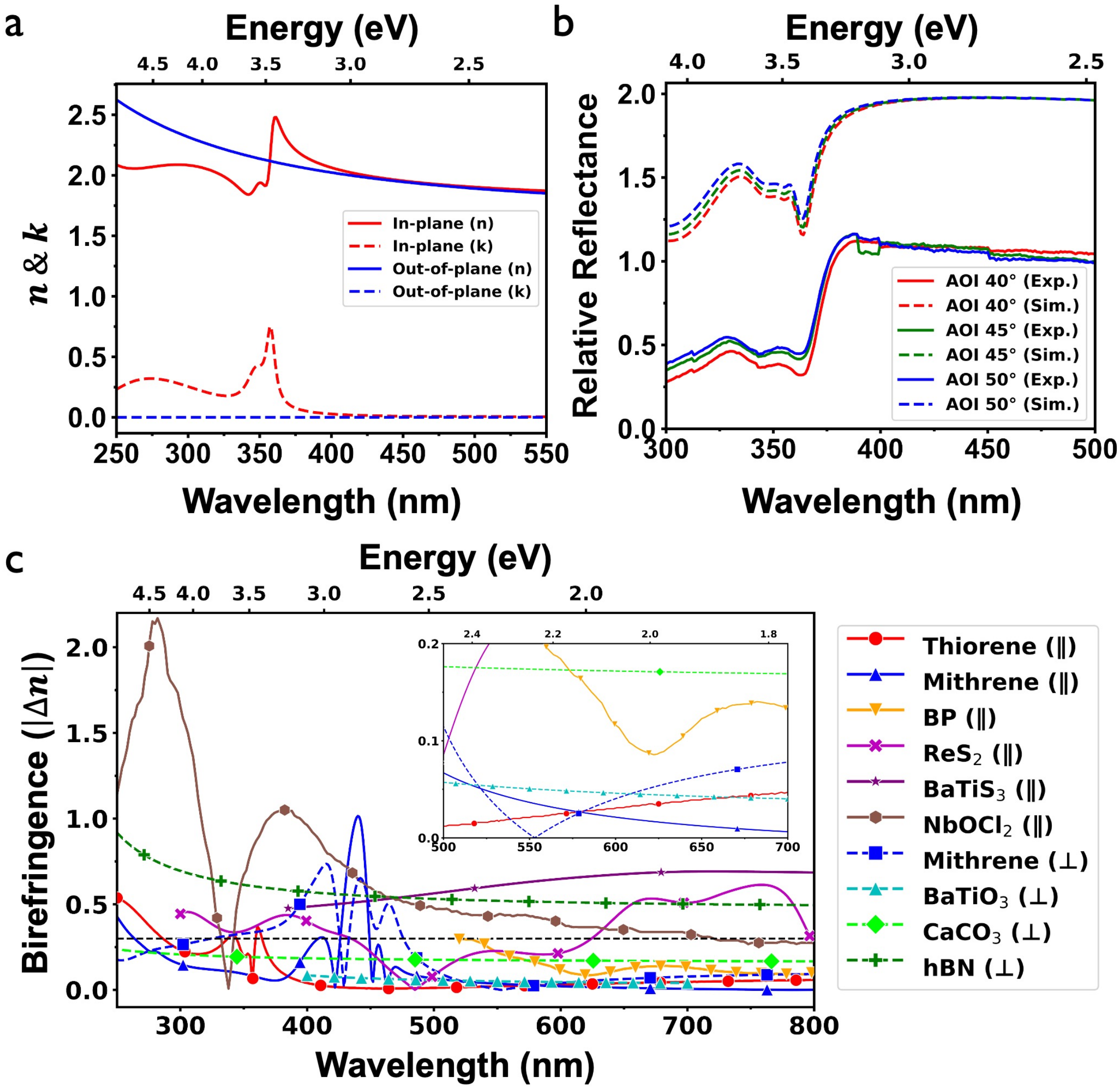

**Figure 2. UV excitons and anisotropic complex refractive indices of thiorene.** (a) In-plane (‖) and out-of-plane (⊥) complex refractive indices ($n$ & $k$) of thiorene extracted from spectroscopic ellipsometry using a uniaxial dielectric-tensor model. A pronounced multiple UV excitonic resonance appears primarily in the in-plane optical response. (b) Angle-resolved reflectance spectra (40°, 45°, and 50°) of a 103 nm thiorene flake on an Aluminum (Al)/ Silicon (Si) substrate, demonstrating good agreement between experiment and transfer matrix method (TMM) simulations, validating the extracted optical constants. (c) Benchmark comparison of birefringence magnitude (|Δn|) across representative anisotropic materials, highlighting the strong UV optical anisotropy of thiorene ($|\Delta n_{UV,avg.}| \sim 0.26$). Note that $|\Delta n|_{\perp}$ represents the birefringence between the in-plane and out-of-plane optical axes, and $|\Delta n|_{\parallel}$

represents the birefringence between two different in-plane optical axes (e.g., $n_a$ vs. $n_b$) in biaxial materials. The horizontal dashed line at |Δn| = 0.3 denotes the commonly adopted threshold for giant birefringence.

Building on the structural/optical identification and dielectric functions obtained from spectroscopic ellipsometry, we next probe the radiative recombination pathways of thiorene through PL spectroscopy. Although thiorene is expected to show PL due to its direct bandgap nature, its wide bandgap (~3.65 eV estimated via the DFT calculation[20]) limits the exploration of PL properties. We excite thiorene using a 248.6 nm (4.99 eV) pulsed laser to generate carriers efficiently above the band edge, which subsequently relax into the lowest bound exciton state. As shown in Fig. 3a, thiorene exhibits strong UV-visible emission centered at ~ 398.8 nm (3.11 eV) with a full width at half maximum (FWHM) of ~ 92.9 nm (0.73 eV) in both bulk (~μm) and thin (~ 20nm) flakes at room-temperature. Although the thin flake exhibits a noisier PL spectrum, likely due to the material's relatively low PL quantum yield (PLQY) and the reduced optical volume, its overall spectral shape and peak position remain nearly identical to the spectrum from the bulk crystal. This spectral consistency despite the reduced thickness confirms the absence of quantum confinement effects, which is expected given the intrinsic MQW structure, similar to 2D Ruddlesden–Popper (RP) perovskite systems[41]. Moreover, PL measurements across multiple flakes show consistent PL emission with minimal flake-to-flake variation (SI Fig. S6). Thiorene's PLQY could potentially be improved through ligand engineering, as demonstrated in MOCs systems[42-44]; such optimization is beyond the scope of this work. We further measure the transmittance of a thin (~20 nm) thiorene flake transferred onto a quartz substrate to exclude possible cavity-induced interference effects. In the transmittance spectrum (dashed red line), a sharp peak at ~3.5 eV is consistent with previously reported diffuse-reflectance absorption[20], which is attributed to the lowest-energy exciton. In addition, a higher-energy feature appears at ~4.5 eV, consistent with the higher-energy resonance identified in ellipsometry. The close agreement between transmittance and ellipsometry confirms the presence of multiple intrinsic excitonic transitions in thiorene. Notably, comparison between the excitonic transmittance dip and emission peak reveals a Stokes shift of approximately 0.41 eV (46.7 nm), indicating that photoexcited excitons relax to lower-energy states before radiative recombination. This large Stokes shift suggests substantial lattice relaxation prior to emission, likely mediated by exciton-phonon interactions in the layered framework, and similar to behavior observed in related MOC systems[27,45].

We further investigate the temperature-dependent PL to probe the evolution of excitonic behavior with temperature. As shown in Fig. 3b, representative spectra at room-temperature (300 K) and low temperature (120 K) reveal that cooling induces a pronounced blue shift of the excitonic peak, accompanied by spectral narrowing. This behavior indicates reduced phonon-assisted broadening and enhanced stabilization of bound excitons at lower temperatures. By fitting the PL spectra using a Lorentz model, we extract both the peak center energy ($E_0$) and the linewidth ($\Gamma$). Lorentzian fit parameter $\Gamma$ denotes the half-width at half-maximum (HWHM); hence, the FWHM is $2\Gamma$. Upon cooling to 120 K, the emission peak blue shifts from 3.11 eV to 3.34 eV, while the FWHM decreases from 92.92 nm (0.73 eV) to 59.21 nm (0.54 eV). Such simultaneous blue shift and linewidth narrowing are characteristic signatures of excitonic materials, reflecting reduced exciton-phonon interactions and suppressed non-radiative scattering channels at low temperature[46]. However, while cooling successfully suppresses thermal broadening, it is noteworthy that the low-temperature

PL emission remains substantially broader than the intrinsic excitonic linewidth extracted from ellipsometry (~60 meV). This large disparity indicates that the emissive state undergoes substantial relaxation and broadening prior to recombination, consistent with the large Stokes shift.

The 2D heat map in Fig. 3c provides a comprehensive visualization of the temperature evolution of the PL emission from 100 K to 300 K in the UV regime. A clear and continuous blue shift of the emission profile is observed with decreasing temperature, accompanied by pronounced spectral narrowing. This behavior is characteristic of excitonic recombination rather than defect-mediated emission. Furthermore, the absence of abrupt emission shifts indicates that no structural phase transitions occur within this temperature range. To quantify this evolution, we extract the PL emission center energy and linewidth using the Lorentzian fitting at each temperature (Fig. 3d). The fitted center energy increases systematically from 3.11 eV at 300 K to 3.53 eV at 100 K, while the Lorentzian linewidth ($\Gamma$, HWHM) decreases from 0.36 eV to 0.24 eV, consistent with bandgap widening and reduced exciton-phonon interactions at lower temperature, reflecting suppression of phonon-assisted scattering. Furthermore, to quantitatively evaluate the exciton-phonon interaction strength in thiorene, we fit the temperature-dependent PL linewidth using the Bose-Einstein type model employed in eq. (1)[22,27]:

$$\Gamma(T) = \Gamma_0 + \frac{\Gamma_{ph}}{\left[exp(\frac{E_{ph}}{k_B T}) - 1\right]} \quad (1)$$

where $\Gamma_0$ represents the temperature-independent scattering rate (i.e., defect and ionized impurities scattering), while $\Gamma_{\mathrm{ph}}$ characterizes the exciton-phonon coupling strength associated with a dominant phonon mode, $k_B$ is the Boltzmann constant, and $\mathrm{E_{ph}}$ is the phonon energy. The characteristic phonon energy contributing to exciton-phonon coupling was taken to be ~100 $cm^{-1}$ (~ 12 meV), guided by prior Raman measurements of mithrene and supported by the presence of a comparable low-energy phonon mode (~100 $cm^{-1}$) in thiorene[27]. The fitting yields $\Gamma_0$ = 215 ± 8 meV and $\Gamma_{\mathrm{ph}}$ = 97 ± 9 meV, with $R^2$ = 0.95, indicating phonon-assisted scattering substantially contributes to the linewidth at elevated temperatures, aligned with the large Stokes shift in thiorene. Note that thiorene exhibits a relatively large $\Gamma_0$, indicating substantial temperature-independent broadening. This residual linewidth likely arises from a combination of static inhomogeneity, defect- or impurity-related scattering, and intrinsic lattice-driven effects such as polaronic dressing associated with its soft lattice, as commonly observed in lead halide perovskites[47,48]. Together, the temperature-dependent spectral shift, linewidth narrowing, and quantitative Lorentz fitting analysis (Fig. 3c, d) confirm that the UV emission in thiorene originates from intrinsic bound excitons that remain robust across a broad temperature range, while exhibiting substantial exciton-phonon coupling.

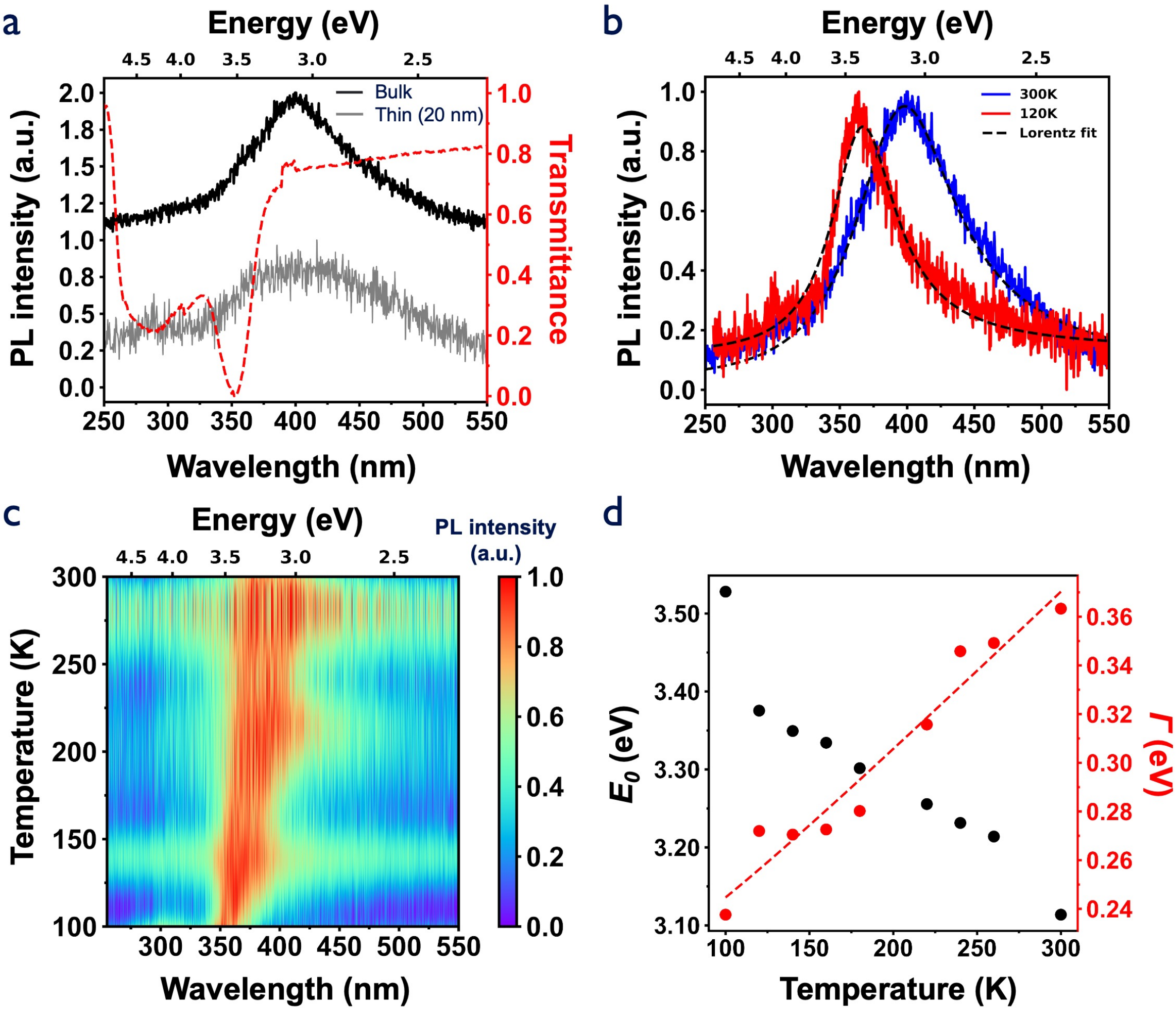

**Figure 3. Temperature-dependent UV photoluminescence (PL) of thiorene.** (a) Transmittance (dashed red) and PL spectra of bulk (~µm) and thin (~20 nm) thiorene flakes. A sharp lowest-energy excitonic resonance (~3.5 eV) and a higher-energy feature (~4.5 eV) are observed, consistent with ellipsometry results. The PL peak at ~3.11 eV exhibits a Stokes shift of ~0.41 eV (~47 nm). (b) Representative PL spectra at 300 K and 120 K. Cooling induces a pronounced blue shift (3.11 → 3.34 eV) and linewidth narrowing (FWHM: 0.73 → 0.54 eV), extracted via Lorentzian fitting. (c) Temperature-dependent PL map (100-300 K) showing continuous blueshift and spectral narrowing with decreasing temperature, characteristic of excitonic bandgap renormalization and phonon-assisted scattering. (d) Temperature dependence of Lorentz-fitted emission parameters. The center energy $E_0$ decreases from 3.53 eV (100 K) to 3.11 eV (300 K), while the linewidth Γ broadens from 0.24 eV to 0.36 eV. The behavior is well described by a Bose-Einstein exciton-phonon coupling model ($R^2$ = 0.95), confirming intrinsic exciton emission with substantial exciton-phonon interaction.

Thiorene exhibits pronounced excitonic resonances and a large refractive index in the UV spectral range, beneficial for the strong light-matter interaction and the formation of EPs. Figure 4a schematically illustrates exciton-photon hybridization, where the coupling between the exciton transition ($E_{\mathrm{exc}}$) and a cavity photon mode ($E_{\mathrm{cav}}$) produces upper and lower EP branches (UEP/LEP) under the strong-coupling condition. To elucidate how cavity architecture impacts the coupling strength for thiorene, we compare two configurations: an open (self-cavity) geometry consisting of thiorene/Aluminum (Al, 100 nm)/Silicon (Si) substrate, and a closed cavity formed by adding a semi-transparent top mirror (thin Al/Thiorene/Al (100 nm)/Si substrate) (Fig.

4a). In the open cavity configuration, the reflective Al substrate together with the relatively high refractive index of thiorene can sustain a lossy Fabry-Pérot optical mode within the thiorene layer itself, enabling self-hybridized EPs without a dedicated top mirror—an approach conceptually aligned with recent demonstrations in other high-index layered excitonic materials[17,49,50], albeit typically with reduced quality factors. On the other hand, the closed cavity configuration provides a tighter cavity mode compared to that of the open system, beneficial for stronger coupling strength. Since the light-matter coupling strength scales as $g \propto \sqrt{f/V_m}$, where $f$ is the oscillator strength and $V_{\mathrm{m}}$ is the mode volume of the cavity, reducing $V_{\mathrm{m}}$ in the closed cavity can increase the coupling strength[1].

Figures 4b and 4c show thickness-dependent reflectance maps (simulation and experiment) for the open and closed cavities, respectively. In both cases, the systems exhibit a clear anticrossing behavior, indicating the formation of UEP and LEP branches. By fitting the branch energies with a two-coupled oscillator model[1], we obtain Rabi splitting of $\hbar\Omega = 488\,meV$ (open cavity) and $\hbar\Omega = 512\,meV$ (closed cavity), respectively, exhibiting an enhanced coupling in the closed cavity geometry. The decay rates of the exciton ($\gamma_{\mathrm{exc}}$) and the cavity mode ($\gamma_{cav}$) are 60 meV and (230/274) meV for the open and closed cavity configurations, respectively. The exciton decay rate is derived from the Lorentzian linewidth obtained through ellipsometry fitting, in Fig. 2a. The extracted coupling strength exceeds the combined dissipation rate ($\hbar\Omega > (\gamma_{cav} + \gamma_{exc})/2$), fulfilling the strong–coupling condition and confirming that thiorene supports robust polariton formation with a large Rabi splitting. The cooperativity, estimated as $C = 2g^2/(\gamma_{\mathrm{exc}}\gamma_{\mathrm{cav}})$, is ~8.6 for the open cavity and ~8.0 for the closed cavity, confirming that coherent exciton–photon coupling dominates over dissipation in both geometries[51]. While the exact exciton relaxation pathways remain to be fully resolved, the large separation between the femtosecond Rabi oscillation period (~ 8 fs) and the picosecond self-trapping time reported in other MOC[52] strongly suggests that polariton formation can precede exciton localization. The Hopfield coefficients extracted from a two-coupled oscillator model quantify the hybrid character: near-zero detuning, both branches exhibit nearly equal exciton and photon fractions, while increasing thickness drives a continuous exchange of photonic and excitonic weights (SI Fig. S7). This hybridization modifies the optical dispersion and can consequently alter the emission characteristics of thiorene, depending on the cavity-exciton detuning, which is controlled by the thickness of the thiorene layer (SI Fig. S8). Notably, the enhanced coupling strength observed in the closed cavity arises from stronger optical confinement. The electric field profiles shown in Fig. 4d confirm the presence of lossy Fabry-Pérot modes in both cavity configurations. The closed-cavity configuration supports a more tightly confined standing-wave profile with thinner thiorene thickness compared to the open cavity configuration, implying a smaller effective mode volume ($V_{\mathrm{m}}$) as expected, enabling the increased coupling strength in the system. Moreover, as the thickness of thiorene increases, multiple cavity mode orders are supported, potentially enabling multimode EPs (SI Fig. S9).

We further investigate the light-matter interaction in thiorene through angle-resolved reflectance measurements, which corroborate the polaritonic nature of the hybrid modes (Fig. 4e). Because the cavity mode exhibits strong angle dependence, the EP hybridization also evolves in an angle-dependent manner. The experimentally extracted dispersion agrees well with the simulated (predicted) EP branches. The cavity mode (red dashed line) displays the expected dispersive behavior with angle, whereas the exciton mode (blue dashed line) remains essentially dispersionless. Finally, we benchmark thiorene against representative EPs platforms such as perovskite, mithrene, GaN, and ZnO in various cavity configurations (open and closed). Fig. 4f highlights thiorene's comparatively large Rabi splitting relative to well-established excitonic materials. Moreover, thiorene exhibits a Rabi splitting that is ~6-10× larger than ZnO and ~3-8× larger than GaN in the UV regime with a simple metallic mirror, even without the use of high-Q DBR cavities typically required in these conventional platforms. This highlights the intrinsically strong light-matter interaction in thiorene. These findings underscore the potential for thiorene as a promising materials platform for UV EPs and polaritonic photochemistry and optoelectronics.

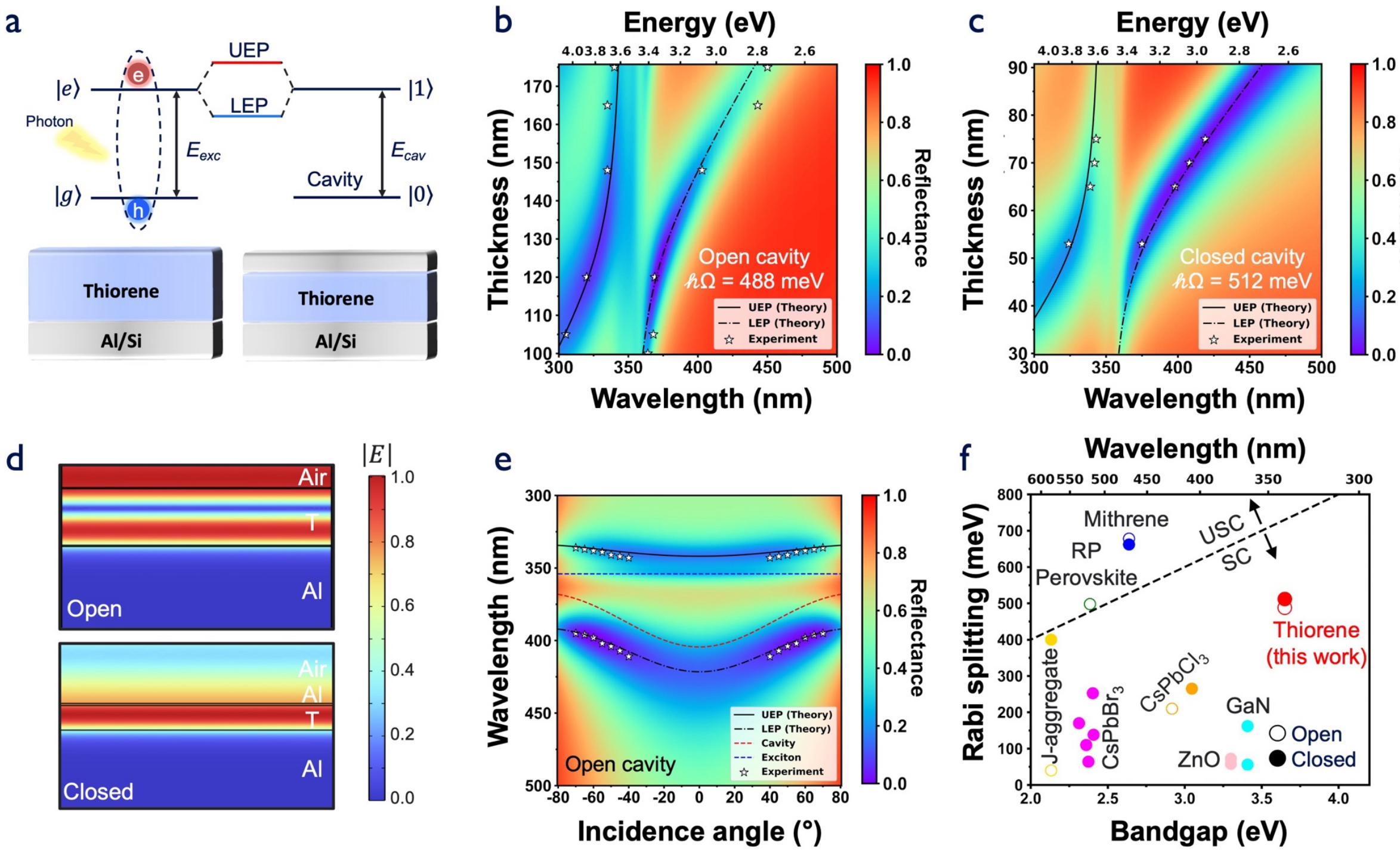

**Figure 4. Strong light–matter interaction in thiorene.** (a) Schematic of exciton–photon coupling in thiorene creating upper and lower exciton-polariton branches (UEP/LEP), and the two cavity configurations studied: open cavity (Thiorene/Al/Si) and closed cavity (Al (~5 nm)/Thiorene/Al/Si). The Al layer thickness (~5 nm) was obtained from TMM fitting and should be regarded as an effective optical thickness rather than a precise physical thickness. (b,c) Thickness-dependent reflectance maps for the open and closed cavities, respectively, showing clear anticrossing between the exciton resonance and Fabry-Pérot cavity mode. The incidence angle was set to 45°, and transverse electric (TE) polarization was used for both simulation and experiment. Solid lines are fits using a two-coupled oscillator model, yielding Rabi splittings of 488 meV (open) and 512 meV (closed). (d) Simulated electric field amplitude profiles for open and closed cavities at λ = 350 nm, indicating stronger field confinement (reduced mode volume) in the closed cavity. The thiorene thickness is set as a zero-detuning thickness. (e) Angle-resolved reflectance and corresponding fitted dispersion (UEP and LEP) confirming polaritonic hybridization in the open cavity system. Red and blue dashed lines indicate the cavity and exciton mode, respectively. (f) Comparison of Rabi splitting versus bandgap for representative EPs platforms (Ruddlesden-Popper (RP) perovskite[53], mithrene[18], $CsPbBr_3$ and $CsPbCl_3$[54-59], J-aggregate[60,61], ZnO[8,9], and GaN[10,11], highlighting the large splitting achieved in thiorene. The dashed line indicates the ultrastrong coupling (USC) condition. Open circles denote open-cavity configurations, while filled circles represent closed-cavity configurations.

## Conclusion

In summary, we establish thiorene as a robust UV excitonic semiconductor and a strong light-matter coupling platform enabled by its intrinsic MQW architecture and pronounced excitonic response. Centimeter-scale single crystals with high compositional uniformity can be mechanically exfoliated into smooth, thickness-controlled flakes, providing a versatile platform for cavity engineering. Spectroscopic ellipsometry reveals a sharp, predominantly in-plane exciton at ~3.46 eV with a narrow linewidth of ~60 meV, while the out-of-plane response remains largely featureless, resulting in strong UV birefringence (~0.3). Temperature-dependent PL further confirms intrinsic excitonic recombination through a characteristic blue shift and linewidth narrowing upon cooling. The UV emission in thiorene is governed by strong exciton-phonon coupling, as evidenced by pronounced temperature-dependent spectral shifts and linewidth evolution, accompanied by a large Stokes shift of 0.41 eV. Leveraging these optical properties, we demonstrate exciton-photon hybridization in both an open self-cavity and a closed metal-mirror cavity with large vacuum Rabi splittings of 488 meV and 512 meV, respectively. The enhanced coupling strength in the closed cavity is consistent with reduced mode volume and stronger field confinement. Together, our results position thiorene as a compelling material system for UV EPs, including dispersion-engineered photonic components, polariton lasers, and polariton-enabled photochemistry. Future efforts to enhance PLQY and optimize cavity design could further improve coherence and enable access to the USC regime in the UV.

## Methods

### Thiorene synthesis

Thiorene crystals were synthesized via an organic single-phase solution method with additional propylamine ($PrNH_2$). Silver nitrate ($AgNO_3$) was dissolved in $PrNH_2$ at a concentration of 5 mM, while diphenyl disulfide ($Ph_2S_2$) was dissolved in toluene at the same concentration. The two precursor solutions were then mixed and stored at ~4 °C for 5–7 days, allowing the growth of large thiorene crystals.

### Sample preparation

The as-synthesized thiorene crystals in $PrNH_2$/Toluene solution were thoroughly washed with isopropyl alcohol (IPA). The cleaned crystals were then drop-cast onto an arbitrary substrate and subsequently picked up using a polydimethylsiloxane (PDMS) stamp for mechanical exfoliation to control the flake thickness. Finally, the exfoliated flakes were dry-transferred onto the target substrate. For the closed-cavity configuration, the top Al was deposited by electron-beam physical vapor deposition (PVDKurt J. Lesker). The thicknesses of exfoliated flakes were determined using tapping mode AFM with a Bruker Dimension Icon system and SCM-PIT-V2 cantilevers (approx. 75 kHz free resonance frequency and 3 N/m force constant).

### Imaging spectroscopic ellipsometry

Imaging spectroscopic ellipsometry measurements were performed using an Accurion EP4 system (Park Systems) over the 250–1000 nm spectral range to determine the dielectric function. A 7× objective lens was employed for measurements in the UV regime. Multi-angle incidence measurements were carried out to improve fitting accuracy. The acquired mapping data were analyzed using the EP4 model software and DataStudio software provided with the Accurion EP4 system.

### Optical characterization (PL, Raman, and Reflectance)

UV-PL spectra were collected using a Photonic Systems DUV Raman/PL 200 spectrometer with 248.6 nm excitation, focused onto the samples with a long working distance 40X objective. For temperature-dependent UV-PL, the crystal was placed inside a liquid nitrogen cooled cell from Linkham, with a UV-transparent $BaF_2$ window. Raman spectra were collected using a Horiba LabRAM HR Evolution confocal microscope equipped with a 600 grooves/mm grating. A 633 nm continuous-wave (CW) laser was used as the excitation source, and the scattered signal was detected with a CCD detector. The angle-resolved reflectance in the UV regime was obtained using the Accurion EP4 system (Park Systems) using ellipsometric contrast microscopy (ECM) mode.

### Optical simulations

Electric field profiles were calculated using finite element method (FEM) simulations performed with the commercial software COMSOL Multiphysics. The theoretical reflectance (normal and oblique incidence) was calculated using the TMM calculation implemented in a Python code adapted from previous literature[62-64]. The refractive index of thiorene was determined through imaging ellipsometry, while that of Al[65] was adopted from previously reported literature values.

### Author Contribution

D.J. supervised and acquired funding for the project. D.J. and Bongjun C. conceived and designed the experiment. Bongjun. C. and Bonnie. C. synthesized the thiorene crystals. Bongjun C. and Bonnie C. performed the ellipsometry measurements and conducted the data fitting. Bongjun C performed the reflectance measurement. T.T.M., R.R., and D.C. (supervised by P.G.) performed the PL spectroscopy. Bongjun C. and M.A.A. performed the AFM measurement. A.A. performed the SEM and EDX measurement. Bongjun C. performed the optical simulations and calculations in discussion with H.-R.K. Bongjun C. and D.J. wrote the manuscript with inputs from all authors. All authors discussed the results and revised the manuscript.

### Acknowledgments

D. J. and B.C. acknowledge support from the Office of Naval Research Young Investigator Award (N00014-23-1-2037). T.T.M. and R.R. acknowledge funding support from the U.S. Air Force Office of Scientific Research LRIR Grant no. 23RXCOR003. P.G. and D.C. acknowledge the support from the Air Force Office of Scientific Research (Grant No. FA9550-22-1-0209). H.-R. K. acknowledges support from the National Research Foundation of Korea (NRF) grant funded by the Korea government (MSIT)(RS-2024-00347397). N.R.G. and M.A. acknowledge funding from the Air Force Office of Scientific Research (AFOSR) under Award No. FA9550-24RYCOR011.

### Competing Interests

The authors declare no competing interests.

Supporting Information

# Ultraviolet Exciton-Polaritons in Silver Phenylthiolate

Bongjun Choi[1], Bonnie Chen[2], Thuc T. Mai[3], Rahul Rao[3], Adam D. Alfieri[1], Du Chen[4,5], Peijun Guo[4,5], Ha-Reem Kim[1,6,7], Michael A. Altvater[3], Nicholas A. Glavin[3], Deep Jariwala[1,2*]

[1]Department of Electrical and Systems Engineering, University of Pennsylvania, Philadelphia, Pennsylvania 19104, United States

[2]Department of Materials Science and Engineering, University of Pennsylvania, Philadelphia, Pennsylvania 19104, United States

[3]Materials and Manufacturing Directorate, Air Force Research Laboratory, Wright-Patterson Air Force Base, OH 45433, USA

[4]Department of Chemical & Environmental Engineering, Yale University, New Haven, CT 06520, USA

[5]Energy Sciences Institute, Yale University, West Haven, CT 06516, USA

[6]Department of Physics, The Catholic University of Korea, Bucheon 14662, Republic of Korea

[7]Department of Physics and Astronomy, and Institute of Applied Physics, Seoul National University, Seoul 08826, Republic of Korea

[*] Corresponding authors: dmj@seas.upenn.edu

**The supporting information file includes:**

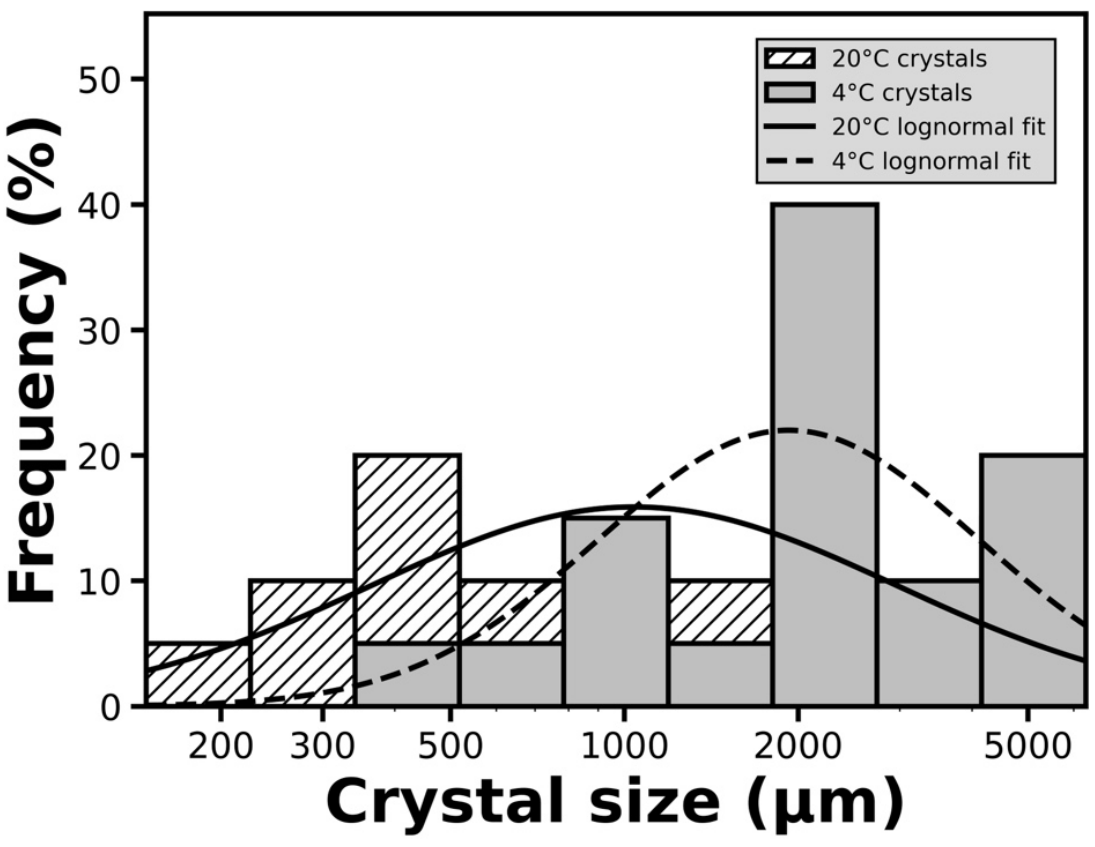

**Figure S1**. **Temperature-dependent crystal size distribution of thiorene (AgSPh).** Crystal size histograms measured after growth at 20°C and 4°C were fitted using lognormal distributions, consistent with the right-skewed nature typical of crystallization-derived size populations. The median crystal size (D50) increased from 1.04 mm at 20 °C to 1.93 mm at 4 °C. The narrower geometric spread at 4 °C (Geometric standard deviation (GSD) = 2.13) relative to 20 °C (GSD = 2.85) indicates more uniform crystal growth under refrigerated conditions. The shift of the distribution peak from approximately $3.48 \times 10^2$ µm at 20°C to $1.09 \times 10^3$ µm at 4°C further supports improved lateral growth and size homogenization at lower temperature.

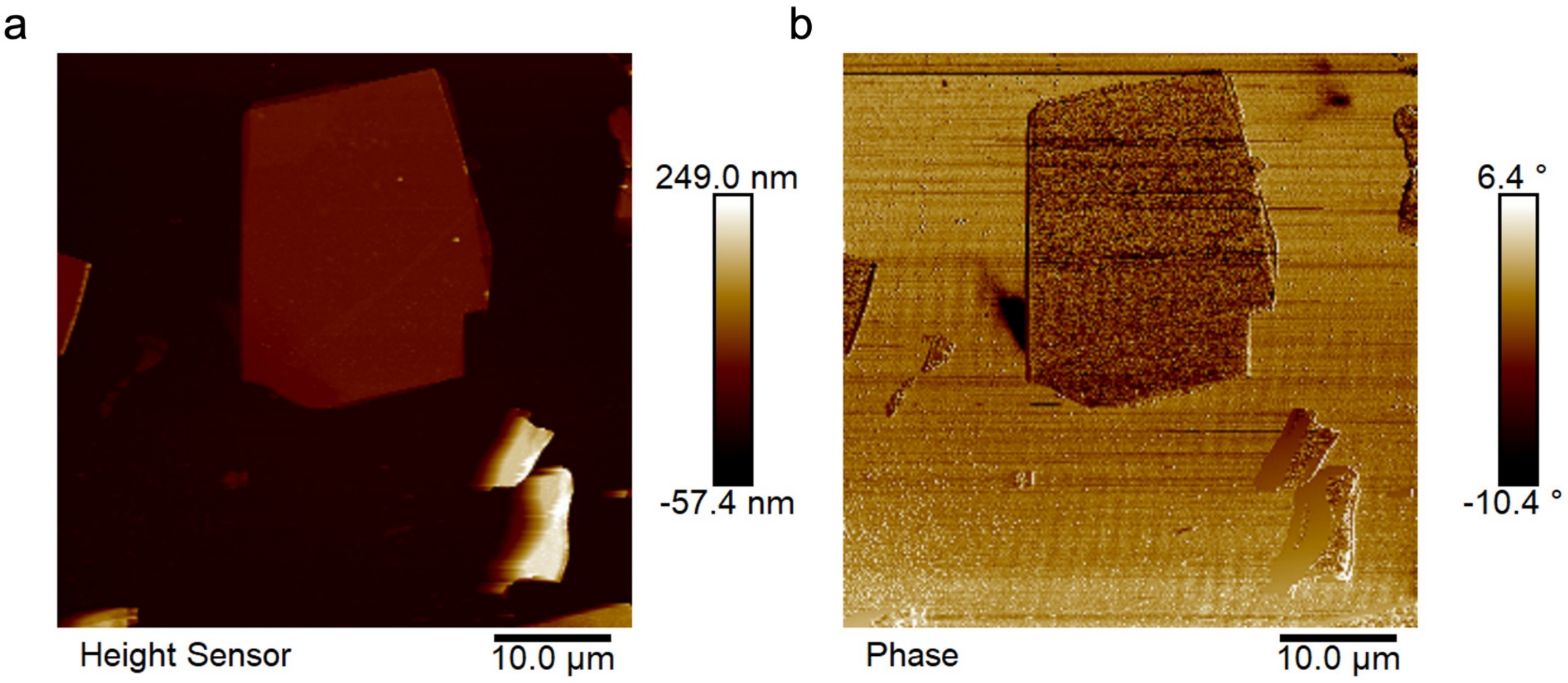

**Figure S2**. **Atomic force microscopy (AFM) images of thiorene.** (a) The height map reveals a smooth and uniform surface morphology. (b) The corresponding phase map exhibits homogeneous contrast without noticeable phase segregation, indicating compositional and structural uniformity across the flake.

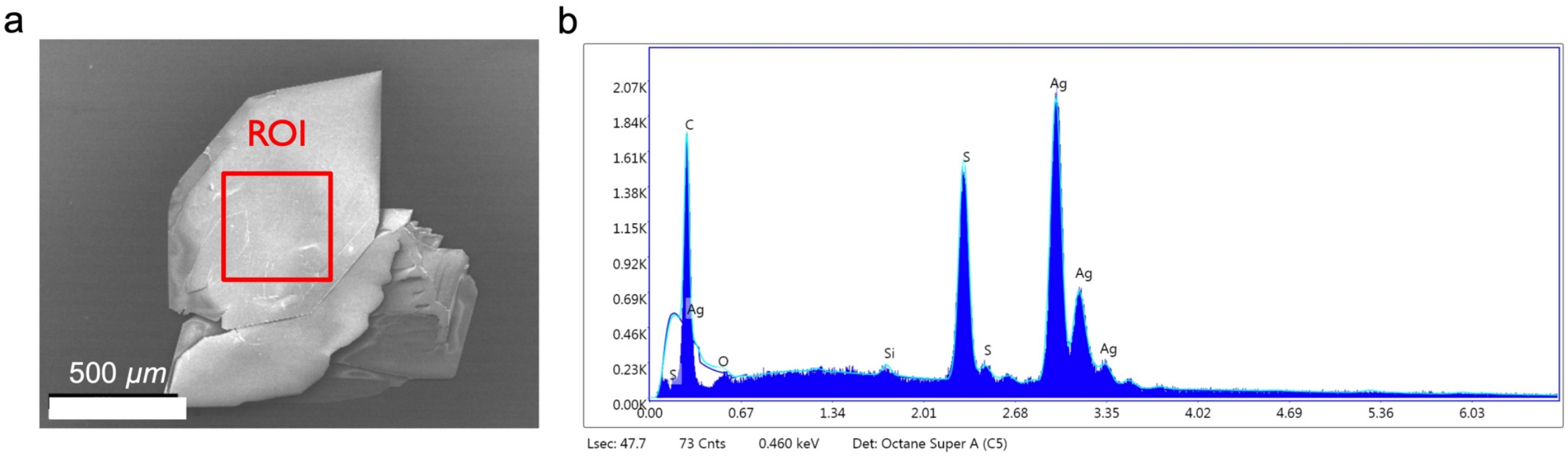

**Figure S3. Energy-dispersive X-ray spectroscopy (EDX) analysis of thiorene.** (a) Scanning electron microscopy (SEM) image of a representative thiorene flake. The red box indicates the region of interest (ROI) selected for elemental analysis. (b) Corresponding EDX spectrum acquired from the ROI, confirming the presence of Ag and S along with C from the phenyl groups, consistent with the expected composition of AgSPh, showing good agreement with the previous literature[1]. The absence of additional impurity peaks indicates high chemical purity of the sample.

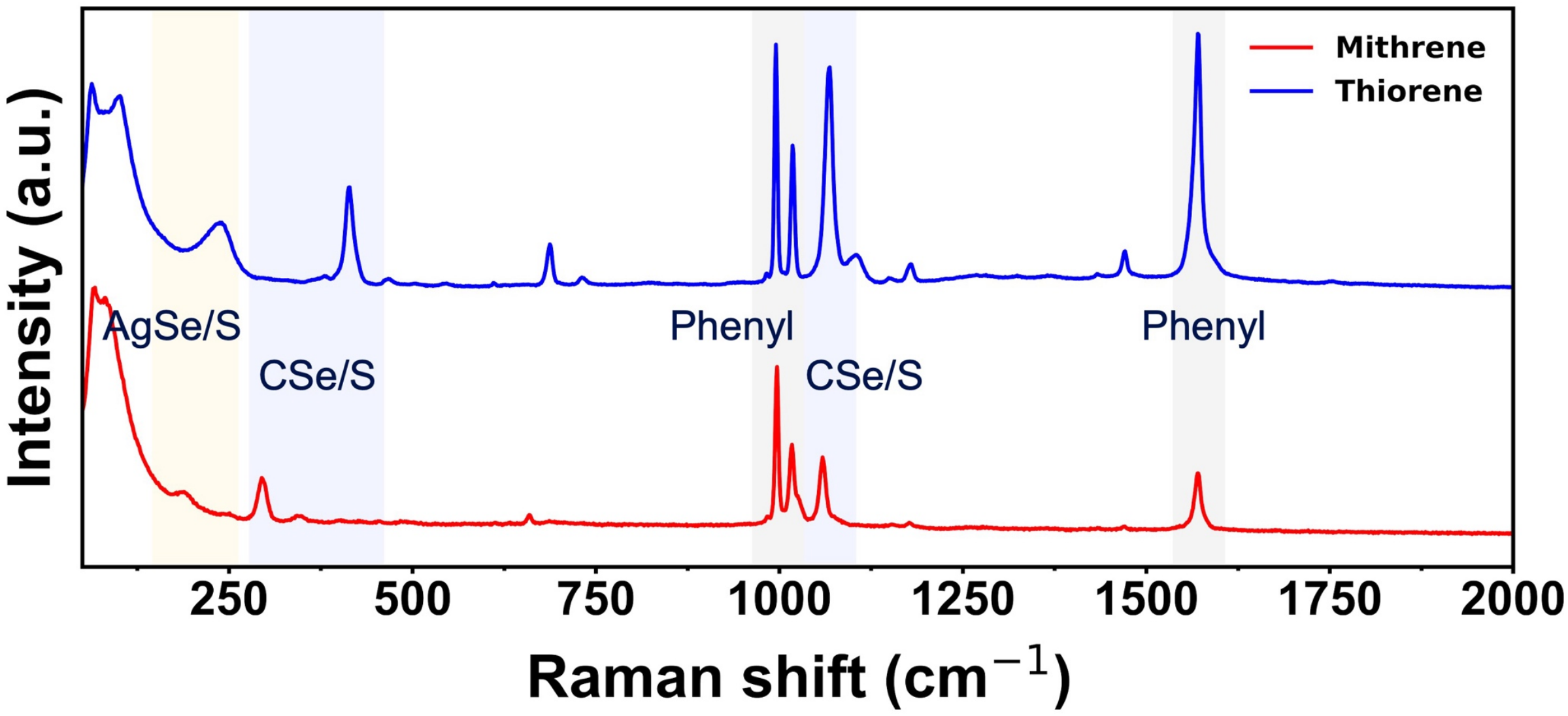

**Figure S4. Raman spectra of mithrene (AgSePh) and thiorene (AgSPh).** Both materials exhibit overall similar spectral features, including characteristic vibrational modes associated with the phenyl rings (grey-shaded region) and prominent low-energy phonon modes (<200 cm$^{-1}$)[2,3]. However, distinct differences in peak positions are observed due to the chalcogen substitution (S vs. Se). In particular, the Ag–S/Se vibrational modes (yellow-shaded region) and the C–S/Se related modes (blue-shaded region) show systematic shifts, reflecting bonding differences between sulfur and selenium.

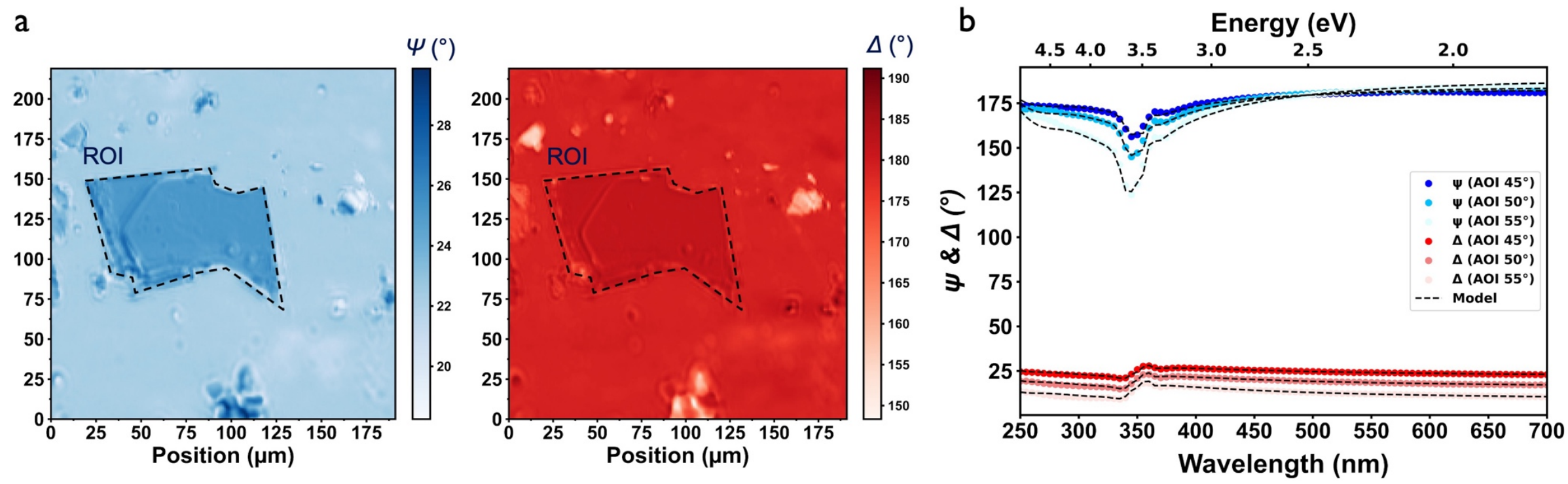

**Figure S5. Imaging spectroscopic ellipsometry analysis of thiorene.** (a) Representative spatial maps of the ellipsometric parameters (Ψ and Δ) obtained from thiorene flakes around the excitonic resonance (λ=280 nm), revealing uniform optical contrast and smooth surface morphology. The ROI was selected for quantitative analysis. (b) Fitting results derived from the extracted ellipsometric spectra. Measurements were performed at multiple angles of incidence to ensure reliable parameter extraction and enhanced sensitivity to anisotropic optical responses. The dielectric function was modeled using a uniaxial model, incorporating multiple Lorentz oscillators for the in-plane response and a Cauchy dispersion model for the out-of-plane direction. This approach reproduces the experimental data with excellent agreement, capturing the anisotropic optical properties of thiorene.

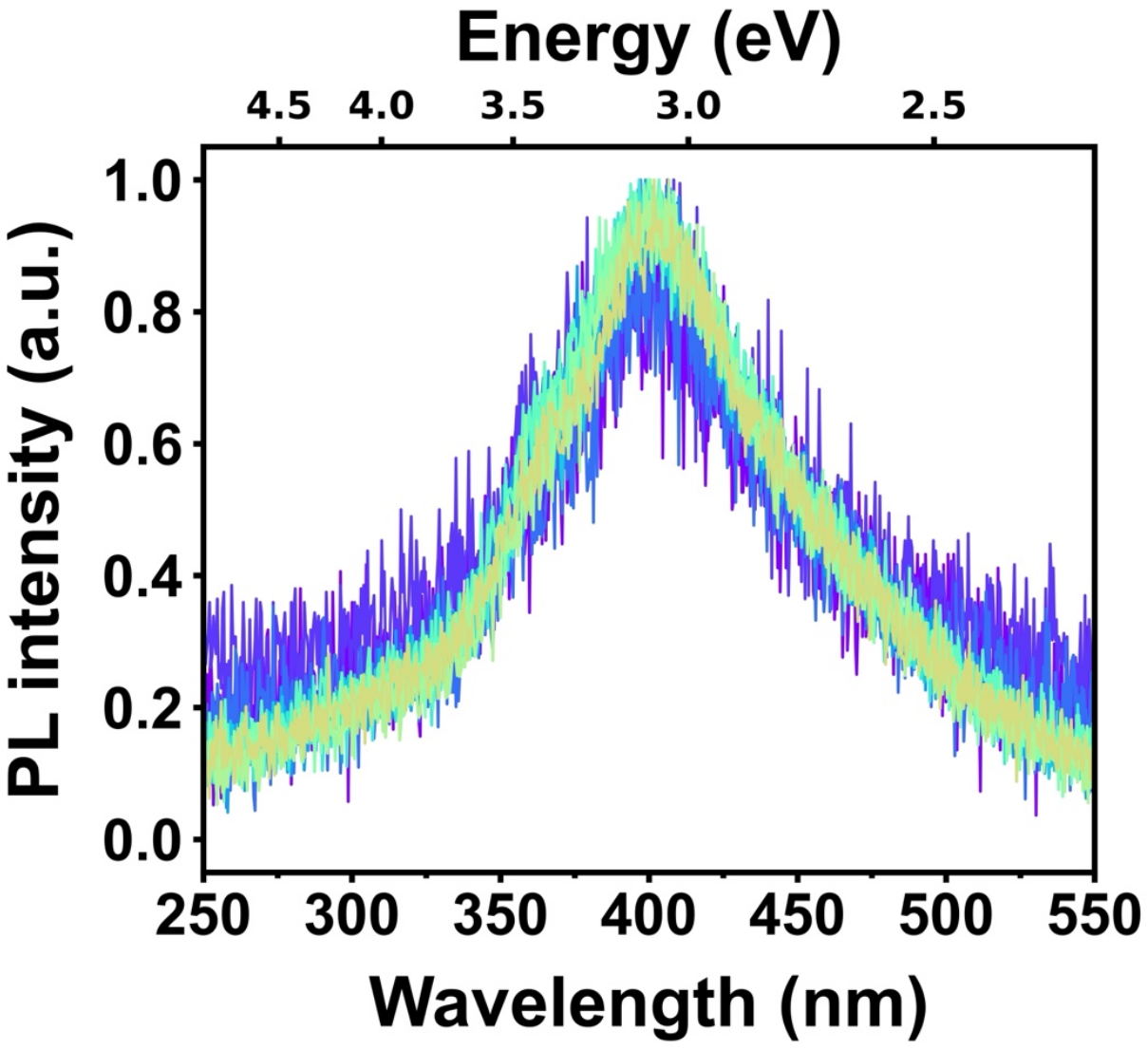

**Figure S6. Photoluminescence (PL) spectra of thiorene.** Acquired from 10 different bulk thiorene flakes, the spectra exhibit nearly identical emission profiles, demonstrating highly reproducible photoluminescence characteristics with minimal flake-to-flake variation.

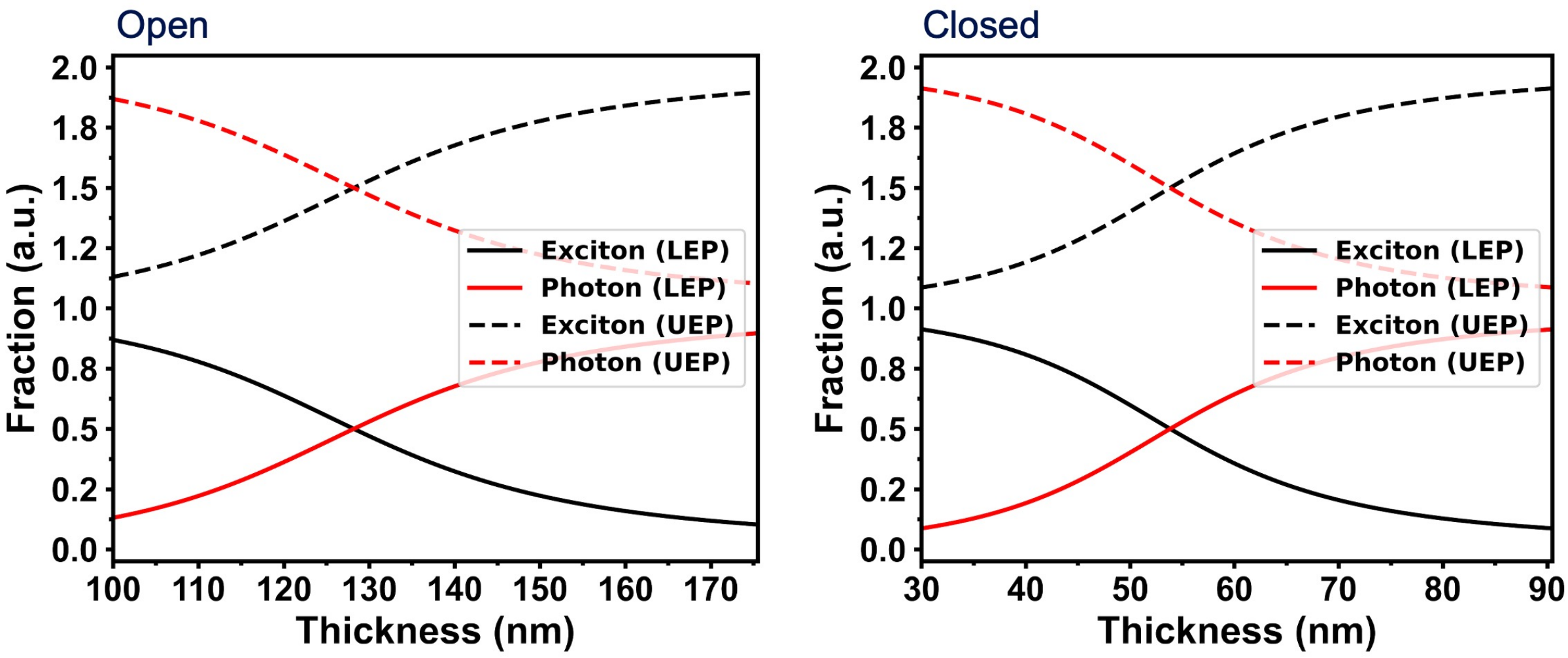

**Figure S7. Thickness-detuned Hopfield coefficients for open and closed cavity configurations.** The excitonic and photonic fractions of the lower exciton–polariton (LEP) and upper exciton–polariton (UEP) branches are plotted as a function of thiorene thickness, extracted using a two-coupled oscillator model[4]. The Hopfield coefficients evolve continuously, revealing the hybrid light–matter character of the polaritonic states.

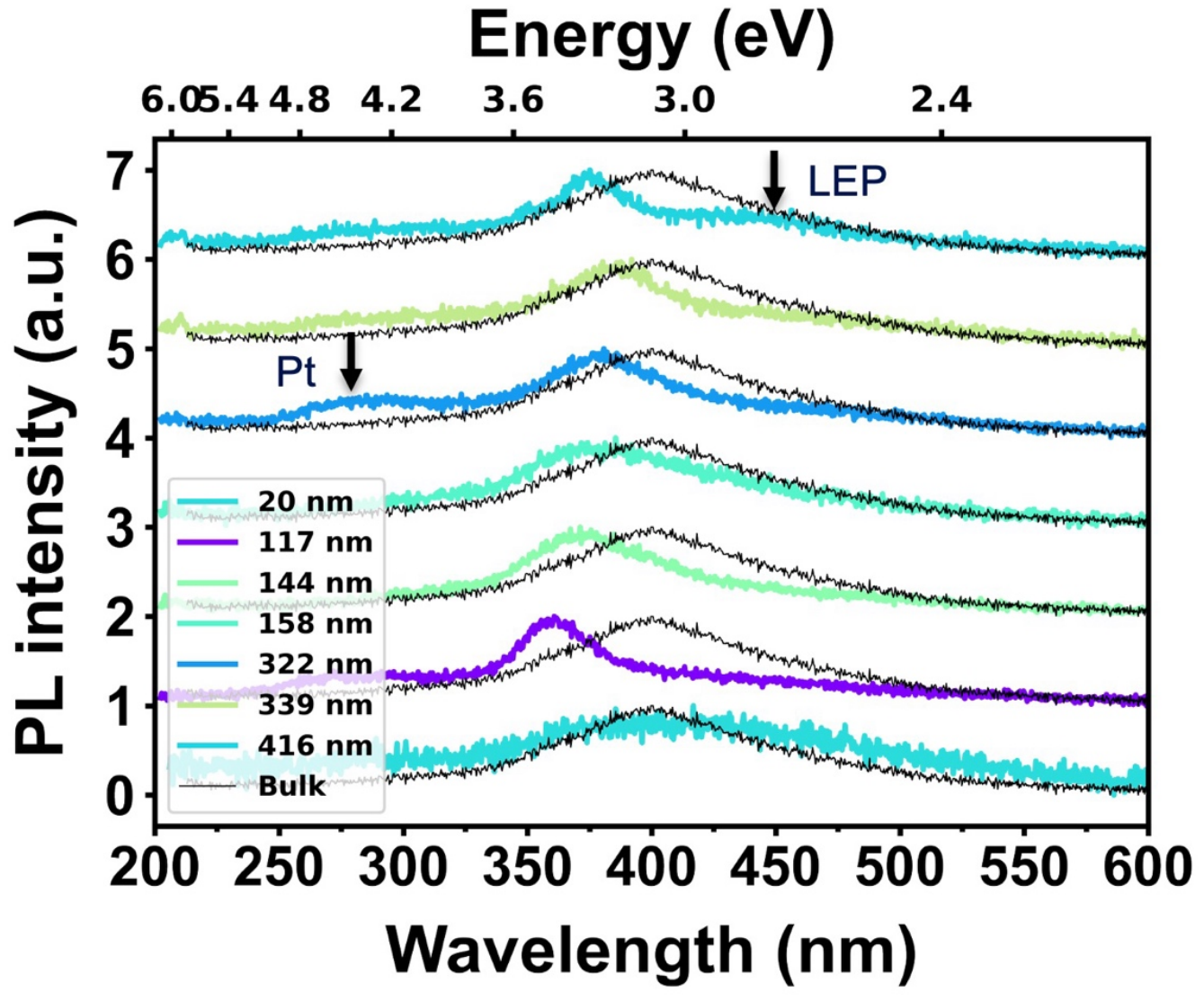

**Figure S8. Thickness-dependent PL of thiorene.** The exfoliated thiroene on Pt substrate samples were used for the measurement. Thin thiorene flakes exhibit emission profiles nearly identical to those of bulk crystals (black lines), indicating negligible cavity effects in this thickness regime since this thin thiorene cannot sustain the self-cavity mode. In contrast, thicker flakes (hundreds of nm), sufficiently thick to sustain a self-cavity mode, display modified emission characteristics arising from exciton–photon hybridization. Specifically, the emission shifts toward ~370 nm, which is attributed to higher-order modes well aligned with the transfer matrix method (TMM) calculation in Figure 4b. A thick flake (~416 nm) further exhibits an additional feature around ~450 nm, likely associated with the LEP branch[3,5]. However, the intrinsically

broad emission of thiorene and its relatively low PL quantum yield (PLQY) limit the clear resolution of distinct polaritonic features. The broad emission observed near 280 nm originates from the Pt substrate.

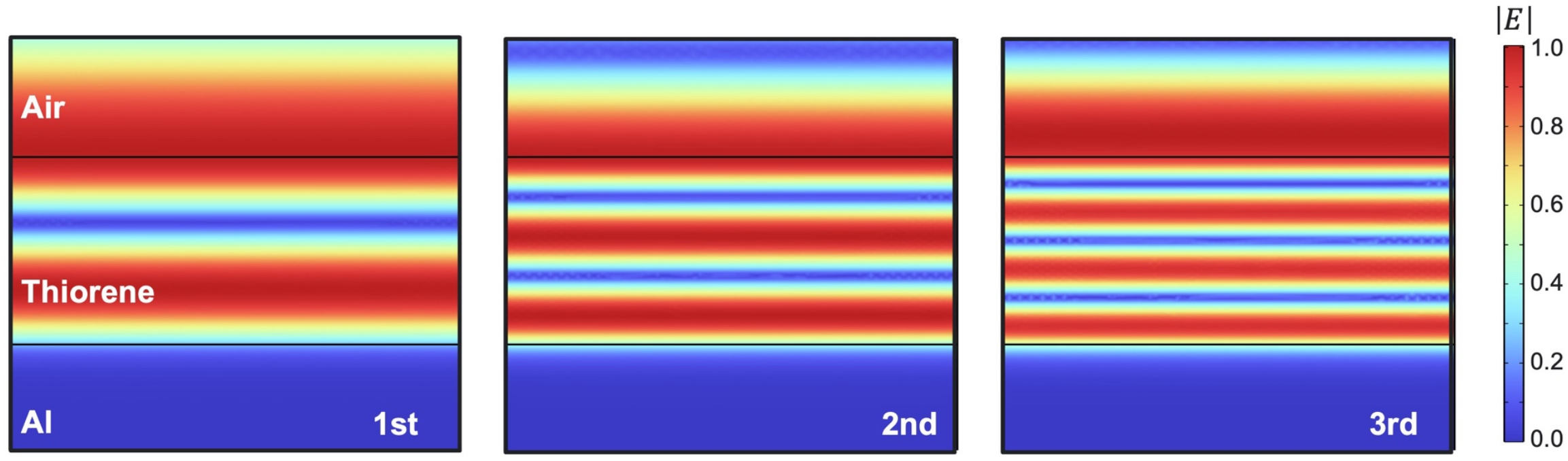

**Figure S9. Electric field distribution in the open cavity configuration.** The electric field profile was simulated using COMSOL Multiphysics for a thiorene (250 nm)/Al structure. The thiorene layer supports multiple lossy Fabry–Pérot resonant modes, including the first-, second-, and third-order modes, which are clearly visible in the spatial electric field distribution. These confined optical modes arise from reflection at the Al interface and partial confinement within the high-index thiorene layer, confirming the formation of self-cavity resonances in the open configuration.

**Table 1. Anisotropic, complex refractive index of thiorene**

| Wavelength (nm) | $n_{in\text{-}plane}$ ($n_o$) | $k_{in\text{-}plane}$ ($k_o$) | $n_{out\text{-}of\text{-}plane}$ ($n_e$) | $k_{out\text{-}of\text{-}plane}$ ($k_e$) |
|---:|---:|---:|---:|---:|
| 250 | 2.088 | 0.23 | 2.625 | 0 |
| 251.875 | 2.078 | 0.243 | 2.61 | 0 |
| 253.75 | 2.07 | 0.256 | 2.595 | 0 |
| 255.625 | 2.064 | 0.267 | 2.58 | 0 |
| 257.5 | 2.06 | 0.278 | 2.566 | 0 |
| 259.375 | 2.057 | 0.287 | 2.552 | 0 |
| 261.25 | 2.056 | 0.295 | 2.539 | 0 |
| 263.125 | 2.056 | 0.302 | 2.526 | 0 |
| 265 | 2.057 | 0.308 | 2.513 | 0 |
| 266.875 | 2.058 | 0.313 | 2.5 | 0 |
| 268.75 | 2.06 | 0.316 | 2.488 | 0 |
| 270.625 | 2.063 | 0.319 | 2.476 | 0 |
| 272.5 | 2.066 | 0.32 | 2.464 | 0 |
| 274.375 | 2.069 | 0.32 | 2.452 | 0 |
| 276.25 | 2.072 | 0.319 | 2.441 | 0 |
| 278.125 | 2.075 | 0.318 | 2.43 | 0 |
| 280 | 2.078 | 0.315 | 2.419 | 0 |
| 281.875 | 2.08 | 0.312 | 2.409 | 0 |
| 283.75 | 2.083 | 0.308 | 2.398 | 0 |
| 285.625 | 2.085 | 0.303 | 2.388 | 0 |
| 287.5 | 2.086 | 0.298 | 2.378 | 0 |
| 289.375 | 2.087 | 0.292 | 2.369 | 0 |
| 291.25 | 2.088 | 0.286 | 2.359 | 0 |
| 293.125 | 2.088 | 0.28 | 2.35 | 0 |
| 295 | 2.088 | 0.273 | 2.341 | 0 |
| 296.875 | 2.087 | 0.266 | 2.332 | 0 |
| 298.75 | 2.085 | 0.259 | 2.323 | 0 |
| 300.625 | 2.083 | 0.252 | 2.314 | 0 |
| 302.5 | 2.08 | 0.244 | 2.306 | 0 |
| 304.375 | 2.077 | 0.237 | 2.298 | 0 |
| 306.25 | 2.073 | 0.23 | 2.289 | 0 |
| 308.125 | 2.068 | 0.223 | 2.281 | 0 |
| 310 | 2.063 | 0.216 | 2.274 | 0 |
| 311.875 | 2.056 | 0.21 | 2.266 | 0 |
| 313.75 | 2.05 | 0.204 | 2.259 | 0 |
| 315.625 | 2.042 | 0.198 | 2.251 | 0 |
| 317.5 | 2.033 | 0.192 | 2.244 | 0 |
| 319.375 | 2.024 | 0.188 | 2.237 | 0 |
| 321.25 | 2.014 | 0.184 | 2.23 | 0 |
| 323.125 | 2.003 | 0.18 | 2.223 | 0 |
| 325 | 1.99 | 0.178 | 2.216 | 0 |
| 326.875 | 1.976 | 0.177 | 2.21 | 0 |
| 328.75 | 1.961 | 0.178 | 2.203 | 0 |
| 330.625 | 1.945 | 0.181 | 2.197 | 0 |
| 332.5 | 1.927 | 0.187 | 2.191 | 0 |
| 334.375 | 1.907 | 0.197 | 2.185 | 0 |
| 336.25 | 1.887 | 0.212 | 2.178 | 0 |
| 338.125 | 1.867 | 0.234 | 2.173 | 0 |
| 340 | 1.85 | 0.266 | 2.167 | 0 |
| 341.875 | 1.841 | 0.307 | 2.161 | 0 |
| 343.75 | 1.848 | 0.354 | 2.155 | 0 |
| 345.625 | 1.873 | 0.397 | 2.15 | 0 |
| 347.5 | 1.909 | 0.421 | 2.144 | 0 |
| 349.375 | 1.936 | 0.427 | 2.139 | 0 |
| 351.25 | 1.937 | 0.433 | 2.134 | 0 |
| 353.125 | 1.917 | 0.476 | 2.129 | 0 |
| 355 | 1.925 | 0.592 | 2.124 | 0 |
| 356.875 | 2.083 | 0.744 | 2.119 | 0 |
| 358.75 | 2.369 | 0.692 | 2.114 | 0 |
| 360.625 | 2.483 | 0.469 | 2.109 | 0 |
| 362.5 | 2.455 | 0.306 | 2.104 | 0 |

| 364.375 | 2.4 | 0.216 | 2.099 | 0 |
|---:|---:|---:|---:|---:|
| 366.25 | 2.35 | 0.164 | 2.095 | 0 |
| 368.125 | 2.309 | 0.132 | 2.09 | 0 |
| 370 | 2.276 | 0.11 | 2.086 | 0 |
| 371.875 | 2.248 | 0.094 | 2.081 | 0 |
| 373.75 | 2.224 | 0.083 | 2.077 | 0 |
| 375.625 | 2.204 | 0.073 | 2.073 | 0 |
| 377.5 | 2.186 | 0.066 | 2.069 | 0 |
| 379.375 | 2.17 | 0.06 | 2.065 | 0 |
| 381.25 | 2.156 | 0.055 | 2.06 | 0 |
| 383.125 | 2.143 | 0.05 | 2.056 | 0 |
| 385 | 2.132 | 0.046 | 2.053 | 0 |
| 386.875 | 2.121 | 0.043 | 2.049 | 0 |
| 388.75 | 2.111 | 0.04 | 2.045 | 0 |
| 390.625 | 2.102 | 0.037 | 2.041 | 0 |
| 392.5 | 2.093 | 0.035 | 2.037 | 0 |
| 394.375 | 2.085 | 0.033 | 2.034 | 0 |
| 396.25 | 2.078 | 0.031 | 2.03 | 0 |
| 398.125 | 2.07 | 0.029 | 2.027 | 0 |
| 400 | 2.064 | 0.027 | 2.023 | 0 |
| 401.875 | 2.057 | 0.026 | 2.02 | 0 |
| 403.75 | 2.051 | 0.024 | 2.016 | 0 |
| 405.625 | 2.045 | 0.023 | 2.013 | 0 |
| 407.5 | 2.04 | 0.022 | 2.01 | 0 |
| 409.375 | 2.034 | 0.021 | 2.006 | 0 |
| 411.25 | 2.029 | 0.02 | 2.003 | 0 |
| 413.125 | 2.024 | 0.019 | 2 | 0 |
| 415 | 2.019 | 0.018 | 1.997 | 0 |
| 416.875 | 2.015 | 0.017 | 1.994 | 0 |
| 418.75 | 2.01 | 0.017 | 1.991 | 0 |
| 420.625 | 2.006 | 0.016 | 1.988 | 0 |
| 422.5 | 2.002 | 0.015 | 1.985 | 0 |
| 424.375 | 1.998 | 0.015 | 1.982 | 0 |
| 426.25 | 1.994 | 0.014 | 1.979 | 0 |
| 428.125 | 1.991 | 0.013 | 1.976 | 0 |
| 430 | 1.987 | 0.013 | 1.973 | 0 |
| 431.875 | 1.984 | 0.013 | 1.971 | 0 |
| 433.75 | 1.98 | 0.012 | 1.968 | 0 |
| 435.625 | 1.977 | 0.012 | 1.965 | 0 |
| 437.5 | 1.974 | 0.011 | 1.962 | 0 |
| 439.375 | 1.971 | 0.011 | 1.96 | 0 |
| 441.25 | 1.968 | 0.011 | 1.957 | 0 |
| 443.125 | 1.965 | 0.01 | 1.955 | 0 |
| 445 | 1.962 | 0.01 | 1.952 | 0 |
| 446.875 | 1.959 | 0.01 | 1.95 | 0 |
| 448.75 | 1.957 | 0.01 | 1.947 | 0 |
| 450.625 | 1.954 | 0.009 | 1.945 | 0 |
| 452.5 | 1.951 | 0.009 | 1.943 | 0 |
| 454.375 | 1.949 | 0.009 | 1.94 | 0 |
| 456.25 | 1.946 | 0.009 | 1.938 | 0 |
| 458.125 | 1.944 | 0.008 | 1.936 | 0 |
| 460 | 1.942 | 0.008 | 1.933 | 0 |
| 461.875 | 1.94 | 0.008 | 1.931 | 0 |
| 463.75 | 1.937 | 0.008 | 1.929 | 0 |
| 465.625 | 1.935 | 0.008 | 1.927 | 0 |
| 467.5 | 1.933 | 0.008 | 1.924 | 0 |
| 469.375 | 1.931 | 0.007 | 1.922 | 0 |
| 471.25 | 1.929 | 0.007 | 1.92 | 0 |
| 473.125 | 1.927 | 0.007 | 1.918 | 0 |
| 475 | 1.925 | 0.007 | 1.916 | 0 |
| 476.875 | 1.923 | 0.007 | 1.914 | 0 |
| 478.75 | 1.921 | 0.007 | 1.912 | 0 |
| 480.625 | 1.92 | 0.007 | 1.91 | 0 |
| 482.5 | 1.918 | 0.007 | 1.908 | 0 |
| 484.375 | 1.916 | 0.007 | 1.906 | 0 |
| 486.25 | 1.914 | 0.007 | 1.904 | 0 |
| 488.125 | 1.913 | 0.006 | 1.902 | 0 |
| 490 | 1.911 | 0.006 | 1.901 | 0 |
| 491.875 | 1.909 | 0.006 | 1.899 | 0 |
| 493.75 | 1.908 | 0.006 | 1.897 | 0 |
| 495.625 | 1.906 | 0.006 | 1.895 | 0 |
| 497.5 | 1.905 | 0.006 | 1.893 | 0 |
| 499.375 | 1.903 | 0.006 | 1.891 | 0 |
| 501.25 | 1.902 | 0.006 | 1.89 | 0 |
| 503.125 | 1.901 | 0.006 | 1.888 | 0 |
| 505 | 1.899 | 0.006 | 1.886 | 0 |
| 506.875 | 1.898 | 0.006 | 1.885 | 0 |
| 508.75 | 1.896 | 0.006 | 1.883 | 0 |
| 510.625 | 1.895 | 0.006 | 1.881 | 0 |
| 512.5 | 1.894 | 0.006 | 1.88 | 0 |
| 514.375 | 1.892 | 0.006 | 1.878 | 0 |
| 516.25 | 1.891 | 0.006 | 1.877 | 0 |
| 518.125 | 1.89 | 0.006 | 1.875 | 0 |
| 520 | 1.889 | 0.006 | 1.873 | 0 |
| 521.875 | 1.888 | 0.005 | 1.872 | 0 |
| 523.75 | 1.886 | 0.005 | 1.87 | 0 |
| 525.625 | 1.885 | 0.005 | 1.869 | 0 |
| 527.5 | 1.884 | 0.005 | 1.867 | 0 |
| 529.375 | 1.883 | 0.005 | 1.866 | 0 |
| 531.25 | 1.882 | 0.005 | 1.864 | 0 |
| 532.1 | 1.881 | 0.005 | 1.864 | 0 |
| 533.125 | 1.881 | 0.005 | 1.863 | 0 |
| 535 | 1.88 | 0.005 | 1.862 | 0 |
| 536.875 | 1.879 | 0.005 | 1.86 | 0 |
| 538.75 | 1.878 | 0.005 | 1.859 | 0 |
| 540.625 | 1.877 | 0.005 | 1.857 | 0 |
| 542.5 | 1.876 | 0.005 | 1.856 | 0 |
| 544.375 | 1.875 | 0.005 | 1.855 | 0 |
| 546.25 | 1.874 | 0.005 | 1.853 | 0 |
| 548.125 | 1.873 | 0.005 | 1.852 | 0 |
| 550 | 1.872 | 0.005 | 1.851 | 0 |
| 551.875 | 1.871 | 0.005 | 1.849 | 0 |
| 553.75 | 1.87 | 0.005 | 1.848 | 0 |
| 555.625 | 1.869 | 0.005 | 1.847 | 0 |
| 557.5 | 1.868 | 0.005 | 1.846 | 0 |
| 559.375 | 1.867 | 0.005 | 1.844 | 0 |
| 561.25 | 1.866 | 0.005 | 1.843 | 0 |
| 563.125 | 1.866 | 0.005 | 1.842 | 0 |
| 565 | 1.865 | 0.005 | 1.841 | 0 |
| 566.875 | 1.864 | 0.005 | 1.839 | 0 |
| 568.75 | 1.863 | 0.005 | 1.838 | 0 |
| 570.625 | 1.862 | 0.005 | 1.837 | 0 |
| 572.5 | 1.861 | 0.005 | 1.836 | 0 |
| 574.375 | 1.861 | 0.005 | 1.835 | 0 |
| 576.25 | 1.86 | 0.005 | 1.834 | 0 |
| 578.125 | 1.859 | 0.005 | 1.833 | 0 |
| 580 | 1.858 | 0.005 | 1.831 | 0 |
| 581.875 | 1.858 | 0.005 | 1.83 | 0 |
| 583.75 | 1.857 | 0.005 | 1.829 | 0 |
| 585.625 | 1.856 | 0.005 | 1.828 | 0 |
| 587.5 | 1.855 | 0.005 | 1.827 | 0 |
| 589.375 | 1.855 | 0.005 | 1.826 | 0 |
| 591.25 | 1.854 | 0.005 | 1.825 | 0 |
| 593.125 | 1.853 | 0.005 | 1.824 | 0 |
| 595 | 1.853 | 0.005 | 1.823 | 0 |
| 596.875 | 1.852 | 0.005 | 1.822 | 0 |

| 598.75 | 1.851 | 0.005 | 1.821 | 0 |
|---|---|---|---|---|
| 600.625 | 1.851 | 0.005 | 1.82 | 0 |
| 602.5 | 1.85 | 0.005 | 1.819 | 0 |
| 604.375 | 1.849 | 0.005 | 1.818 | 0 |
| 606.25 | 1.849 | 0.005 | 1.817 | 0 |
| 608.125 | 1.848 | 0.005 | 1.816 | 0 |
| 610 | 1.848 | 0.005 | 1.815 | 0 |
| 611.875 | 1.847 | 0.005 | 1.814 | 0 |
| 613.75 | 1.846 | 0.005 | 1.813 | 0 |
| 615.625 | 1.846 | 0.005 | 1.812 | 0 |
| 617.5 | 1.845 | 0.005 | 1.811 | 0 |
| 619.375 | 1.845 | 0.005 | 1.81 | 0 |
| 621.25 | 1.844 | 0.005 | 1.809 | 0 |
| 623.125 | 1.843 | 0.005 | 1.809 | 0 |
| 625 | 1.843 | 0.005 | 1.808 | 0 |
| 626.875 | 1.842 | 0.005 | 1.807 | 0 |
| 628.75 | 1.842 | 0.005 | 1.806 | 0 |
| 630.625 | 1.841 | 0.005 | 1.805 | 0 |
| 632.5 | 1.841 | 0.005 | 1.804 | 0 |
| 634.375 | 1.84 | 0.005 | 1.803 | 0 |
| 636.25 | 1.84 | 0.005 | 1.803 | 0 |
| 638.125 | 1.839 | 0.005 | 1.802 | 0 |
| 640 | 1.839 | 0.005 | 1.801 | 0 |
| 641.875 | 1.838 | 0.004 | 1.8 | 0 |
| 643.75 | 1.838 | 0.004 | 1.799 | 0 |
| 645.625 | 1.837 | 0.004 | 1.798 | 0 |
| 647.5 | 1.837 | 0.004 | 1.798 | 0 |
| 649.375 | 1.836 | 0.004 | 1.797 | 0 |
| 651.25 | 1.836 | 0.004 | 1.796 | 0 |
| 653.125 | 1.835 | 0.004 | 1.795 | 0 |
| 655 | 1.835 | 0.004 | 1.795 | 0 |
| 656.875 | 1.834 | 0.004 | 1.794 | 0 |
| 658.75 | 1.834 | 0.004 | 1.793 | 0 |
| 660.625 | 1.833 | 0.004 | 1.792 | 0 |
| 662.5 | 1.833 | 0.004 | 1.792 | 0 |
| 664.375 | 1.832 | 0.004 | 1.791 | 0 |
| 666.25 | 1.832 | 0.004 | 1.79 | 0 |
| 668.125 | 1.832 | 0.004 | 1.789 | 0 |
| 670 | 1.831 | 0.004 | 1.789 | 0 |
| 671.875 | 1.831 | 0.004 | 1.788 | 0 |
| 673.75 | 1.83 | 0.004 | 1.787 | 0 |
| 675.625 | 1.83 | 0.004 | 1.787 | 0 |

| 677.5 | 1.829 | 0.004 | 1.786 | 0 |
|---|---|---|---|---|
| 679.375 | 1.829 | 0.004 | 1.785 | 0 |
| 681.25 | 1.829 | 0.004 | 1.784 | 0 |
| 683.125 | 1.828 | 0.004 | 1.784 | 0 |
| 685 | 1.828 | 0.004 | 1.783 | 0 |
| 686.875 | 1.827 | 0.004 | 1.782 | 0 |
| 688.75 | 1.827 | 0.004 | 1.782 | 0 |
| 690.625 | 1.827 | 0.004 | 1.781 | 0 |
| 692.5 | 1.826 | 0.004 | 1.781 | 0 |
| 694.375 | 1.826 | 0.004 | 1.78 | 0 |
| 696.25 | 1.826 | 0.004 | 1.779 | 0 |
| 698.125 | 1.825 | 0.004 | 1.779 | 0 |
| 700 | 1.825 | 0.004 | 1.778 | 0 |
| 701.875 | 1.824 | 0.004 | 1.777 | 0 |
| 703.75 | 1.824 | 0.004 | 1.777 | 0 |
| 705.625 | 1.824 | 0.004 | 1.776 | 0 |
| 707.5 | 1.823 | 0.004 | 1.776 | 0 |
| 709.375 | 1.823 | 0.004 | 1.775 | 0 |
| 711.25 | 1.823 | 0.004 | 1.774 | 0 |
| 713.125 | 1.822 | 0.004 | 1.774 | 0 |
| 715 | 1.822 | 0.004 | 1.773 | 0 |
| 716.875 | 1.822 | 0.004 | 1.773 | 0 |
| 718.75 | 1.821 | 0.004 | 1.772 | 0 |
| 720.625 | 1.821 | 0.004 | 1.771 | 0 |
| 722.5 | 1.821 | 0.004 | 1.771 | 0 |
| 724.375 | 1.82 | 0.004 | 1.77 | 0 |
| 726.25 | 1.82 | 0.004 | 1.77 | 0 |
| 728.125 | 1.82 | 0.004 | 1.769 | 0 |
| 730 | 1.819 | 0.004 | 1.769 | 0 |
| 731.875 | 1.819 | 0.004 | 1.768 | 0 |
| 733.75 | 1.819 | 0.004 | 1.767 | 0 |
| 735.625 | 1.818 | 0.004 | 1.767 | 0 |
| 737.5 | 1.818 | 0.004 | 1.766 | 0 |
| 739.375 | 1.818 | 0.004 | 1.766 | 0 |
| 741.25 | 1.818 | 0.004 | 1.765 | 0 |
| 743.125 | 1.817 | 0.004 | 1.765 | 0 |
| 745 | 1.817 | 0.004 | 1.764 | 0 |
| 746.875 | 1.817 | 0.004 | 1.764 | 0 |
| 748.75 | 1.816 | 0.004 | 1.763 | 0 |
| 750.625 | 1.816 | 0.004 | 1.763 | 0 |
| 752.5 | 1.816 | 0.004 | 1.762 | 0 |
| 754.375 | 1.816 | 0.004 | 1.762 | 0 |

| 756.25 | 1.815 | 0.004 | 1.761 | 0 |
|---|---|---|---|---|
| 758.125 | 1.815 | 0.004 | 1.761 | 0 |
| 760 | 1.815 | 0.004 | 1.76 | 0 |
| 761.875 | 1.814 | 0.004 | 1.76 | 0 |
| 763.75 | 1.814 | 0.004 | 1.759 | 0 |
| 765.625 | 1.814 | 0.004 | 1.759 | 0 |
| 767.5 | 1.814 | 0.004 | 1.758 | 0 |
| 769.375 | 1.813 | 0.004 | 1.758 | 0 |
| 771.25 | 1.813 | 0.004 | 1.757 | 0 |
| 773.125 | 1.813 | 0.004 | 1.757 | 0 |
| 775 | 1.813 | 0.004 | 1.757 | 0 |
| 776.875 | 1.812 | 0.004 | 1.756 | 0 |
| 778.75 | 1.812 | 0.004 | 1.756 | 0 |
| 780.625 | 1.812 | 0.004 | 1.755 | 0 |
| 782.5 | 1.812 | 0.004 | 1.755 | 0 |
| 784.375 | 1.811 | 0.004 | 1.754 | 0 |
| 786.25 | 1.811 | 0.004 | 1.754 | 0 |
| 788.125 | 1.811 | 0.004 | 1.753 | 0 |
| 790 | 1.811 | 0.004 | 1.753 | 0 |
| 791.875 | 1.81 | 0.004 | 1.753 | 0 |
| 793.75 | 1.81 | 0.004 | 1.752 | 0 |
| 795.625 | 1.81 | 0.004 | 1.752 | 0 |
| 797.5 | 1.81 | 0.004 | 1.751 | 0 |
| 799.375 | 1.809 | 0.004 | 1.751 | 0 |
| 801.25 | 1.809 | 0.004 | 1.75 | 0 |
| 803.125 | 1.809 | 0.004 | 1.75 | 0 |
| 805 | 1.809 | 0.004 | 1.75 | 0 |
| 806.875 | 1.809 | 0.004 | 1.749 | 0 |
| 808.75 | 1.808 | 0.004 | 1.749 | 0 |
| 810.625 | 1.808 | 0.004 | 1.748 | 0 |
| 812.5 | 1.808 | 0.004 | 1.748 | 0 |
| 814.375 | 1.808 | 0.004 | 1.748 | 0 |
| 816.25 | 1.807 | 0.004 | 1.747 | 0 |
| 818.125 | 1.807 | 0.004 | 1.747 | 0 |
| 820 | 1.807 | 0.004 | 1.746 | 0 |
| 821.875 | 1.807 | 0.004 | 1.746 | 0 |
| 823.75 | 1.807 | 0.004 | 1.746 | 0 |
| 825.625 | 1.806 | 0.004 | 1.745 | 0 |
| 827.5 | 1.806 | 0.004 | 1.745 | 0 |
| 829.375 | 1.806 | 0.004 | 1.744 | 0 |
| 831.25 | 1.806 | 0.004 | 1.744 | 0 |
| 833.125 | 1.806 | 0.004 | 1.744 | 0 |

| 835 | 1.805 | 0.004 | 1.743 | 0 |
|---|---|---|---|---|
| 836.875 | 1.805 | 0.004 | 1.743 | 0 |
| 838.75 | 1.805 | 0.004 | 1.743 | 0 |
| 840.625 | 1.805 | 0.004 | 1.742 | 0 |
| 842.5 | 1.805 | 0.004 | 1.742 | 0 |
| 844.375 | 1.804 | 0.004 | 1.742 | 0 |
| 846.25 | 1.804 | 0.004 | 1.741 | 0 |
| 848.125 | 1.804 | 0.004 | 1.741 | 0 |
| 850 | 1.804 | 0.004 | 1.741 | 0 |
| 851.875 | 1.804 | 0.004 | 1.74 | 0 |
| 853.75 | 1.803 | 0.004 | 1.74 | 0 |
| 855.625 | 1.803 | 0.004 | 1.739 | 0 |
| 857.5 | 1.803 | 0.004 | 1.739 | 0 |
| 859.375 | 1.803 | 0.004 | 1.739 | 0 |
| 861.25 | 1.803 | 0.004 | 1.738 | 0 |
| 863.125 | 1.803 | 0.004 | 1.738 | 0 |
| 865 | 1.802 | 0.004 | 1.738 | 0 |
| 866.875 | 1.802 | 0.004 | 1.737 | 0 |
| 868.75 | 1.802 | 0.004 | 1.737 | 0 |
| 870.625 | 1.802 | 0.004 | 1.737 | 0 |
| 872.5 | 1.802 | 0.004 | 1.737 | 0 |
| 874.375 | 1.802 | 0.004 | 1.736 | 0 |
| 876.25 | 1.801 | 0.004 | 1.736 | 0 |
| 878.125 | 1.801 | 0.004 | 1.736 | 0 |
| 880 | 1.801 | 0.004 | 1.735 | 0 |
| 881.875 | 1.801 | 0.004 | 1.735 | 0 |
| 883.75 | 1.801 | 0.004 | 1.735 | 0 |
| 885.625 | 1.801 | 0.004 | 1.734 | 0 |
| 887.5 | 1.8 | 0.004 | 1.734 | 0 |
| 889.375 | 1.8 | 0.004 | 1.734 | 0 |
| 891.25 | 1.8 | 0.004 | 1.733 | 0 |
| 893.125 | 1.8 | 0.004 | 1.733 | 0 |
| 895 | 1.8 | 0.004 | 1.733 | 0 |
| 896.875 | 1.8 | 0.004 | 1.733 | 0 |
| 898.75 | 1.799 | 0.004 | 1.732 | 0 |
| 900.625 | 1.799 | 0.004 | 1.732 | 0 |
| 902.5 | 1.799 | 0.004 | 1.732 | 0 |
| 904.375 | 1.799 | 0.004 | 1.731 | 0 |
| 906.25 | 1.799 | 0.004 | 1.731 | 0 |
| 908.125 | 1.799 | 0.004 | 1.731 | 0 |
| 910 | 1.799 | 0.004 | 1.731 | 0 |
| 911.875 | 1.798 | 0.004 | 1.73 | 0 |

| 913.75 | 1.798 | 0.004 | 1.73 | 0 |
|---|---|---|---|---|
| 915.625 | 1.798 | 0.004 | 1.73 | 0 |
| 917.5 | 1.798 | 0.004 | 1.729 | 0 |
| 919.375 | 1.798 | 0.004 | 1.729 | 0 |
| 921.25 | 1.798 | 0.004 | 1.729 | 0 |
| 923.125 | 1.798 | 0.004 | 1.729 | 0 |
| 925 | 1.797 | 0.004 | 1.728 | 0 |
| 926.875 | 1.797 | 0.004 | 1.728 | 0 |
| 928.75 | 1.797 | 0.004 | 1.728 | 0 |
| 930.625 | 1.797 | 0.004 | 1.728 | 0 |
| 932.5 | 1.797 | 0.004 | 1.727 | 0 |
| 934.375 | 1.797 | 0.004 | 1.727 | 0 |
| 936.25 | 1.797 | 0.004 | 1.727 | 0 |
| 938.125 | 1.796 | 0.004 | 1.726 | 0 |
| 940 | 1.796 | 0.004 | 1.726 | 0 |
| 941.875 | 1.796 | 0.004 | 1.726 | 0 |
| 943.75 | 1.796 | 0.004 | 1.726 | 0 |
| 945.625 | 1.796 | 0.004 | 1.725 | 0 |
| 947.5 | 1.796 | 0.004 | 1.725 | 0 |
| 949.375 | 1.796 | 0.004 | 1.725 | 0 |
| 951.25 | 1.795 | 0.004 | 1.725 | 0 |
| 953.125 | 1.795 | 0.004 | 1.724 | 0 |
| 955 | 1.795 | 0.004 | 1.724 | 0 |
| 956.875 | 1.795 | 0.004 | 1.724 | 0 |
| 958.75 | 1.795 | 0.004 | 1.724 | 0 |
| 960.625 | 1.795 | 0.004 | 1.723 | 0 |
| 962.5 | 1.795 | 0.004 | 1.723 | 0 |
| 964.375 | 1.795 | 0.004 | 1.723 | 0 |
| 966.25 | 1.794 | 0.004 | 1.723 | 0 |
| 968.125 | 1.794 | 0.004 | 1.723 | 0 |
| 970 | 1.794 | 0.004 | 1.722 | 0 |
| 971.875 | 1.794 | 0.004 | 1.722 | 0 |
| 973.75 | 1.794 | 0.004 | 1.722 | 0 |
| 975.625 | 1.794 | 0.004 | 1.722 | 0 |
| 977.5 | 1.794 | 0.004 | 1.721 | 0 |
| 979.375 | 1.794 | 0.004 | 1.721 | 0 |
| 981.25 | 1.794 | 0.004 | 1.721 | 0 |
| 983.125 | 1.793 | 0.004 | 1.721 | 0 |
| 985 | 1.793 | 0.004 | 1.72 | 0 |
| 986.875 | 1.793 | 0.004 | 1.72 | 0 |
| 988.75 | 1.793 | 0.004 | 1.72 | 0 |
| 990.625 | 1.793 | 0.004 | 1.72 | 0 |
| 992.5 | 1.793 | 0.004 | 1.72 | 0 |
| 994.375 | 1.793 | 0.004 | 1.719 | 0 |
| 996.25 | 1.793 | 0.004 | 1.719 | 0 |
| 998.125 | 1.793 | 0.004 | 1.719 | 0 |
| 1000 | 1.792 | 0.004 | 1.719 | 0 |